\documentclass[12pt]{article}

\usepackage{amsmath, amssymb, amsfonts, amsthm}
\usepackage{graphicx}
\usepackage{natbib}
\usepackage{indentfirst}
\usepackage{times}
\usepackage{bbm}
\usepackage{lineno}
\usepackage{authblk}
\usepackage{afterpage}

\usepackage[colorlinks, linkcolor=black, citecolor=black,filecolor=black, urlcolor=black]{hyperref}

\newcommand{\deriv}{\ensuremath{d}}

\makeatletter

\newcounter{id}
\renewcommand{\theid}{I\arabic{id}}
\renewcommand{\theHid}{I\arabic{id}}

\newcommand{\c@org@eq}{}
\let\c@org@eq\c@equation
\newcommand{\org@theeq}{}
\let\org@theeq\theequation

\newcommand{\setid}{
\let\c@equation\c@id
\let\theHequation\theHid
\let\theequation\theid}

\newcommand{\setreg}{
\let\c@equation\c@org@eq
\let\theHequation\org@theHeq
\let\theequation\org@theeq}

\makeatother

\begin{document}

\title{Formal properties of the probability of fixation: identities, inequalities and approximations}

\author[1]{David M.~McCandlish\thanks{To whom correspondence should be addressed. E-mail: \texttt{davidmc@sas.upenn.edu}}}
\author[2]{Charles L.~Epstein}
\author[1]{Joshua B.~Plotkin}
\affil[1]{Department of Biology, University of Pennsylvania, Philadelphia, PA}
\affil[2]{Department of Mathematics, University of Pennsylvania, Philadelphia, PA}

\date{}






{\let\newpage\relax\maketitle}

\begin{center}
 \vspace{1cm}
\end{center}

\begin{abstract}  \normalsize 
\noindent 
The formula for the probability of fixation of a new mutation is widely used in theoretical population genetics and molecular evolution. Here we
derive a series of identities, inequalities and approximations for the exact probability of fixation of a new mutation under the Moran process
(equivalent results hold for the approximate probability of fixation under the Wright-Fisher process, after an appropriate change of variables). We
show that the logarithm of the fixation probability has a particularly simple behavior when the selection coefficient is measured as a difference
of Malthusian fitnesses, and we exploit this simplicity to derive inequalities and approximations. We also present a comprehensive
comparison of both existing and new approximations for the fixation probability, highlighting those approximations that induce a reversible Markov
chain when used to describe the dynamics of evolution under weak mutation.
\end{abstract}

\vfill


\setlength{\parindent}{.62cm}
\setlength{\parskip}{2ex plus0.5ex minus0.2ex}

\setid
\setreg

\section{Introduction}

A basic goal of evolutionary biology is to understand the process by which a new mutant allele, present in only a single copy within a population, can
eventually come to be carried by all members of a population. This process is somewhat difficult to understand in general, both because of
its dependence on the vicissitudes of the lives and deaths of particular individual organisms and because of the vast diversity of ways that this
process could unfold (i.e.~the enormous set of possible paths through the space of gene frequencies). It is therefore useful to consider
certain simple summary statistics capable of providing broad insights into this complex process~\citep[e.g.][]{Ewens04}. One productive
approach is to abstract away from the time-dependent aspects of this process and to consider only the final outcome, i.e.~to consider the
probability that a new mutant allele is destined to fix in the population or, conversely, that the new allele is destined to be
lost~\citep{Fisher22,Haldane27,Fisher30,Fisher30b,Wright31,Malecot52,Kimura57,Kimura62}. Formulae for this probability are typically
expressed as functions of both the magnitude of natural selection (differences in expected number of offspring, etc.) and the form of genetic
drift (population size, distributions of offspring number, etc.). Such formulae therefore provide a quantitative understanding of the relationship between
these key factors in the evolutionary process.

Besides providing a framework for considering the relative influence of natural selection and genetic drift in evolution, formulae for the
probability of fixation play a widespread role as constituent elements in more complex models of evolution~\citep[reviewed in][]{McCandlish13f}. The
most common way of constructing such models is to assume that mutation is sufficiently weak that each new mutation is either lost or fixed before
the next new mutation enters the population; under this assumption one can construct a model of evolution in which the fate of each new mutation is based solely on its probability of fixation, either by assuming that new mutations enter the population with some distribution of selection coefficients and are fixed or lost independently from each other~\citep{Ohta77,Kimura79,Sawyer92}, or by considering a sequence of fixation events at a single locus~\citep{Gillespie83, Iwasa88, Bulmer91, Hartl98, Orr02,Sella05}. 

In light of the importance of the probability of fixation for understanding the relationship between selection and drift and for contemporary
models of evolution under weak mutation, it is unfortunate that the standard formulae for the fixation probability are, mathematically, difficult
to understand and manipulate. As a result of these difficulties, various approximations to the probability of fixation have been proposed and are in
common use. However, these approximations suffer from a number of defects, such as being uncontrolled (no bounds on the direction and magnitude of the
error) and altering fundamental aspects of the evolutionary dynamics~\citep[e.g., making deleterious substitutions impossible,][]{McCandlish13e}. It
is therefore useful to study the probability of fixation in more detail, with the dual goals of better understanding the relationship between selection and drift and constructing better approximations.

Here, we develop a series of identities, inequalities and approximations for the exact probability of fixation of a new mutation under a Moran process~\citep{Moran59}. This probability is given by
\begin{equation}
\label{eq:def}
u_N(s) = \frac{1-e^{-s}}{1-e^{-Ns}}.
\end{equation}
where $N$ is the population size and $s$ is the difference between the $\log$ fitness of the invading type and the $\log$ fitness of the resident type (i.e.~the difference in Malthusian fitness). Defining the selection coefficient as the difference in log fitnesses is useful because it puts the probability of fixation into a form very similar to that of the approximate probability of fixation under a Wright-Fisher process~\citep{Kimura57,Kimura62}. This means that the results described here can easily be translated into approximate results for the Wright-Fisher process under an appropriate rescaling of the variables. In fact, many of our inequalities can be extended to provide exact bounds on the Wright-Fisher fixation probability (see Discussion).

Our approach to studying the behavior of the probability of fixation stems from the following remarkable (albeit well-known) identity, which
relates the fixation probability of mutations whose selection coefficients have equal magnitude but opposite signs:
\begin{equation}
\label{eq:wellknown}
\frac{u_{N}(s)}{u_{N}(-s)}=e^{(N-1)s}.
\end{equation}
This identity plays a particularly important role in multi-allelic models of evolution where it is assumed that each new mutation is lost or fixed before the next new mutation enters the population. In such ``sequential fixations'' models, evolution is formalized as a Markov chain on the set of alleles (rather than the set of allele frequencies), and the transition rate between allele $i$ and allele $j$ is given by
\begin{equation}
Q(i,j)=N\,\mu_{i,j}\, u_{N} (s_{i,j}) \quad \text{for } i \neq j
\end{equation}
where $\mu_{i,j}$ is the mutation rate from $i$ to $j$ and $s_{i,j}$ is the difference between the $\log$ fitness of allele $j$ and the $\log$ fitness of allele $i$. Under mild additional assumptions, Equation~\ref{eq:wellknown} determines the equilibrium distribution of this sequential fixations Markov chain and ensures that it is reversible~\citep{Iwasa88,Sella05,Berg04,Manhart12}. Here, we use this identity as the basis for providing a simpler and more intuitive analysis of the effects of the selection coefficient on the probability of fixation and for constructing new approximations that maintain the reversibility and equilibrium distribution of these sequential fixations Markov chains.

One immediate consequence of the above identity is that the logarithm of the probability of fixation must obey certain symmetries and so it may be
easier to understand than the probability of fixation itself. Indeed, we show that the $\log$ probability of fixation is concave, and we use this
$\log$-concavity to establish a number of useful inequalities. In addition, our results on the $\log$ probability of fixation provide deeper insights
into the relationship between selection and drift than studying the probability of fixation itself. This is because taking $\log$s allows us to ask
how a small change in selection coefficient increases or decreases the probability of fixation in a proportional, rather than absolute, sense.

Another consequence of this identity is that given an approximation for the probability of fixation for advantageous mutations one can construct a natural extension of that approximation for deleterious mutations by stipulating that the approximation should also satisfy the identity. Any approximation constructed in this fashion automatically preserves the structure of the equilibrium distribution of the corresponding sequential fixations Markov chain and the reversibility of the resulting dynamics. We provide a comprehensive comparison between existing approximations and a set of new approximations derived by this method, paying close attention to approximations that also serve as upper or lower bounds on the probability of fixation. Notably, we find that the approximation
\begin{equation}
Nu_{N}(s)\approx \left\{\begin{aligned} & S+ e^{-S/2} && \text{for } S > 0\\
  &-S\,e^{S}+e^{3S/2} && \text{for } S\leq0.
\end{aligned} \right.
\end{equation}
performs quite well compared to the standard approximation~\citep{Fisher30,Wright31}
\begin{equation}
\label{eq:introWF}
Nu_{N}(s)\approx \frac{S}{1-e^{-S}},
\end{equation}
where $S=Ns$, and it should provide a useful substitute in many circumstances (the main advantage of the new approximation is that it is not
expressed as a fraction, making it easier to treat as an integrand). We demonstrate the utility of this new approximation by showing how it simplifies the calculation of the rate of evolution under several common choices for the distribution of mutational effects on fitness~\citep{Ohta77,Kimura79,Piganeau03}.

In the processes of developing these ideas, we highlight many other results that give additional insight into the probability of fixation of a new
mutation. For instance, we show that the common approximation in Equation~\ref{eq:introWF} has the surprising feature that the probability of
fixation can be additively partitioned into one term, corresponding to the probability of fixation in an infinite population and another term,
capturing the effects of finite population-size, which depends only on the magnitude of the selection coefficient and not its sign. We explore the implications of this decomposition in the Discussion.

It is worth noting that while we conduct a thorough analysis of the exact probability of fixation of a new mutation for the Moran model under
frequency-independent selection, there are many other aspects of the probability of fixation that we make no attempt to cover~\citep[see][for a comprehensive review]{Patwa08}. For an analysis of the monotonicity of the probability of fixation starting from an arbitrary frequency and under arbitrary diploid selection, see~\citep{Chen08}; for frequency-dependent selection, see~\citep{Taylor04,Wu13}; for the accuracy of the approximate probability of fixation under the diffusion approximation, see~\citep{Moran60,Burger95}; for the probability of fixation in structured populations see~\citep{Whitlock03} and for the more general case of evolution in graph-structured populations, see~\citep{Nowak06,Shakarian12}; recent results for time-varying selection and population size can be found in~\citep{Uecker11,Waxman11,Peischl12}. 

\section{Basic properties of the probability of fixation}

The probability of fixation for a new mutation under the Moran process, $u_{N}(s)$, was given in Equation~\ref{eq:def}. In what follows, we will assume $N>1$, since if $N=1$ then $u_N(s)=1$ for all $s$. Furthermore, for notational convenience, we will generally leave the index $N$ in $u_{N}(s)$ implicit and simply write $u(s)$ in places where doing so introduces no ambiguity.

Some of the behavior of $u(s)$ is easy to determine based on its functional form, and thus some of its behavior is readily apparent. It is easy to see that $0<u(s)<1$ and one can use L'H\^{o}pital's rule twice to show that $u(0)=1/N$. Further simplifications are possible if we assume that certain terms are large or small. In particular, for large, positive $Ns$ the probability of fixation behaves like $1-e^{-s}$, which for small $s$ behaves like $s$; for large, negative $s$ the probability of fixation behaves like $e^{(N-1)s}$; and for large, negative $Ns$ but small $s$, the probability of fixation behaves like $-s\,e^{Ns}$.

\setid

\begin{table}[p]
\begin{align}
\label{eq:exp} &\frac{u(s)}{u(-s)} =e^{(N-1)s}  & &   \text{Reversibility} \\
\label{eq:inv} &\frac{1}{u(s)} = \sum_{k=0}^{N-1}e^{-ks}  & &   \text{Finite geometric sum} \\
\label{eq:rec} &u_{N}(s) =\frac{u_{N-1}(s)}{e^{-s}+u_{N-1}(s)} & & \text{Recursive formula} \\
\label{eq:almostsym} &u(s)-(1-e^{-s})=e^{-s}\,u(-s) & & \text{Comparison with $Ns \gg1$ limit}\\
&u(s) =e^{(N-1)s/2}\left(\frac{\sinh (s/2)}{\sinh (Ns/2)}\right) & &  \text{Hyperbolic identity} \\
\label{eq:hyp} &u(s)u(-s) =\left(\frac{\sinh (s/2)}{\sinh (Ns/2)}\right)^{2} & & \text{Hyperbolic identity II} \\
\label{eq:derivforineq} &u'(s) =1-u(s)-e^{-s}u'(-s) && \text{Derivative with respect to $s$} 
\end{align}
\caption{\label{tab:identities}Basic identities for the probability of fixation. The above identities are all easy to verify. Equation~\ref{eq:exp} is well-known; equation~\ref{eq:inv} can be found in~\citep{Traulsen07}.} 
\end{table}
\afterpage{\clearpage}

\setreg

Other features of the probability of fixation require a little more work to derive. For convenience, we have included a list of useful, general
identities in Table~\ref{tab:identities}. In what follows, we will also require a basic understanding about the derivatives of $u(s)$.

Although intuitively it is obvious that the probability of fixation should be increasing in $s$, the fact that $u'(s)$ is always positive is not immediately apparent from the functional form of $u(s)$ or from direct evaluation of $u'(s)$~\citep[for a broader study of when the probability of fixation behaves monotonically as a function of either selection coefficients or initial frequency, see][]{Chen08}. However, differentiating both sides of Equation~\ref{eq:inv} from Table~\ref{tab:identities} and rearranging gives us:
\begin{equation}
u'(s)=u(s)^{2}\sum_{k=1}^{N-1}k\,e^{-ks}
\end{equation}
which is clearly positive for all $s$. Furthermore, using the fact that $u'(s)>0$ for all $s$, Equation~\ref{eq:derivforineq} from
Table~\ref{tab:identities} gives us $u'(s)<1$, so that indeed both $0<u(s)<1$ and $0<u'(s)<1$. The fact that $u'(s)<1$ is biologically
interesting because it is common to approximate $u(s)\approx s$ for large $Ns$ and small $s$; this approximation thus always overestimates the sensitivity of $u(s)$ to changes in $s$.

Unlike $u'(s)$, little can be said about the behavior of $u''(s)$, at least analytically. Intuitively, $u''(s)$ must be negative for large, positive $s$ (where $u(s)\approx 1-e^{-s}$) and positive for large, negative $s$ (where $u(s)\approx e^{(N-1)s}$). Numerical exploration suggests that the corresponding point of inflection in $u(s)$ occurs at $s \approx (\log N)/N$. 

As we have seen, it is possible to make some limited progress determining the features of the probability of fixation by working directly with $u(s)$. One of the main messages in what follows is that the situation simplifies considerably if we consider the $\log$ probability of fixation instead.

\section{Identities: the logarithmic approach}

There are two reasons for consider the $\log$ probability of fixation. The first reason, which is more important biologically, is that the derivative
$\frac{\deriv}{\deriv s}\log u(s)$ provides, in some sense, a more informative measure of how changing $s$ changes the probability of fixation, than
$u'(s)$ itself does. This is because $\frac{\deriv}{\deriv s}\log u(s)=u'(s)/u(s)$, i.e.~the derivative of the $\log$ probability of fixation describes how changes in $s$ affect the probability of fixation in a relative, rather than absolute sense. For instance, making a deleterious fixation slightly more fit may hardly increase its probability of fixation when measured in terms of a difference of probabilities, but may nonetheless cause a substantial proportional increase in the probability of fixation. If most mutations are strongly deleterious, making such mutations slightly less deleterious might have a large impact on the rate of evolution, an effect that would be completely hidden if we only considered $u'(s)$.

The second reason to study the $\log$ probability of fixation is one we have already mentioned: the behavior of the $\log$ probability of fixation is much simpler than the behavior of the probability of fixation. This simplicity arises in large part due to Equation~\ref{eq:exp} in Table~\ref{tab:identities}, and can be seen by taking the natural logarithm of both sides of that equation. It is not immediately obvious, but this simple behavior is also closely related to the fact that the sequential fixations Markov chain is often reversible.

To see the relationship between the simple behavior of the $\log$ probability of fixation and the reversibility of the sequential fixations Markov chain, let us step back and consider the probability of fixation as being just some function $q_{N}$ with $0<q_{N}(s)<1$. Suppressing the subscript $N$ in our notation as for $u(s)$, the corresponding sequential fixations Markov chain has the rate matrix
\begin{equation}
Q(i,j)=N\,\mu_{i,j}\, q( \log f_{j}-\log f_{i}) \quad \text{for } i \neq j
\end{equation}
where $f_{i}$ is the fitness of allele $i$, $\mu_{i,j}$ is the mutation rate from $i$ to $j$, and the diagonal entries of $Q$ are specified by the requirement that the row sums be zero. Now, let us assume that there are at least three alleles and that the $\mu_{i,j}$ are chosen such that the sequential fixations Markov chain is reversible for the neutral case, i.e. $q(s)=1/N$. One might wonder what form the probability of fixation must take such that adding arbitrary frequency-independent natural selection to this model preserves its reversibility. It turns out that a necessary and sufficient condition is that $q$ must satisfy the identity:
\begin{equation}
\frac{q(s)}{q(-s)} \times \frac{q(\tilde{s})}{q(-\tilde{s})}=\frac{q(s+\tilde{s})}{q(-(s+\tilde{s}))}
\end{equation}
for all $s$ and $\tilde{s}$, which in turn implies that for all $s$
\begin{equation}
\label{eq:reversableQ}
\frac{q(s)}{q(-s)}=e^{\nu(N)\, s}
\end{equation}  
for some function $\nu$ that does not depend on $s$~\citep{Manhart12}. Taking $\log$s and then differentiating, we have:
\begin{gather}
\log q(s)=\nu(N)\,s+\log q(-s) \label{eq:linerar} \\ 
\frac{\deriv}{\deriv s} \log q(s)=\nu(N)+\frac{\deriv}{\deriv s}\log q(-s)\label{eq:derivlog}\\
\frac{\deriv^{2}}{\deriv s^{s}} \log q(s)=\frac{\deriv^{2}}{\deriv s^{2}}\log q(-s) \label{eq:even}
\end{gather}
assuming, of course, that the relevant derivatives are defined.
Thus, we see immediately that, for any probability of fixation that results in the reversibility of the sequential fixations Markov chain, the second derivative of the $\log$ probability of fixation with respect to fitness is an even function, while the first derivative is equal to $\nu(N)/2$ plus some odd function of $s$ (recall that $f$ is known as an even function if $f(x)=f(-x)$ for all $x$; $f$ is odd if $f(x)=-f(-x)$ for all $x$).

\begin{figure}
\center
\includegraphics[width=10cm]{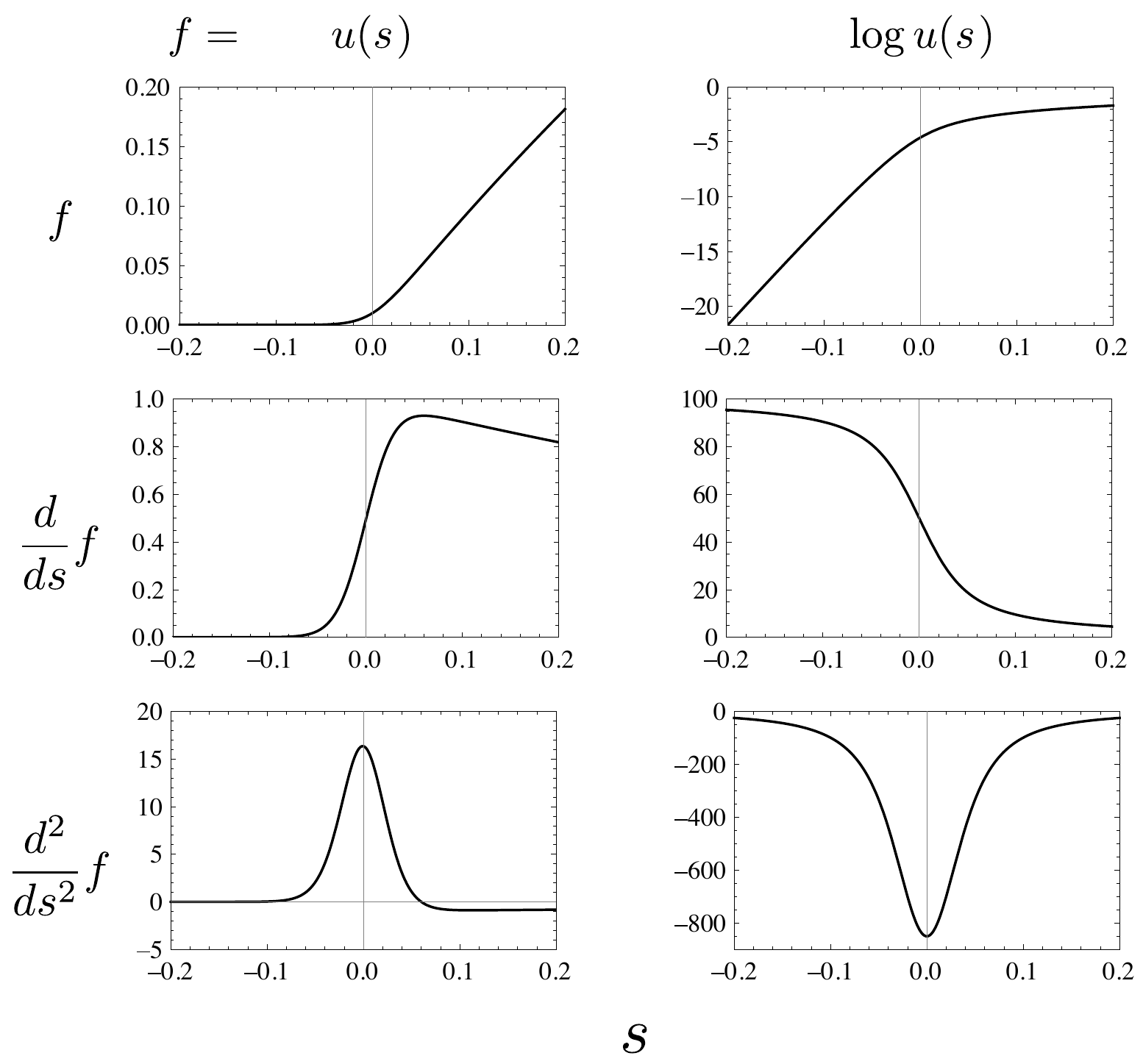}
\caption[]{
A comparison between (left column) the probability of fixation and its derivatives with respect to $s$ and (right column) the $\log$ probability of fixation and its derivates with respect to $s$. $N=101$.
}
\label{fig:comparison}
\end{figure}

Returning to the actual probability of fixation for a Moran process, $u(s)$, by Equation~\ref{eq:exp} in Table~\ref{tab:identities} we have $\nu(N)=N-1$. Figure~\ref{fig:comparison} compares $u(s)$ and its derivatives with respect to $s$ to $\log u(s)$ and its corresponding derivatives. The figure suggests that the derivatives of $\log u(s)$ have several useful features even beyond those guaranteed by the above considerations. For instance, unlike the first derivative of $u(s)$ with respect to $s$, the first derivative of $\log u(s)$ with respect to $s$ changes monotonically in $s$. It is this simplicity that will allow us to develop many of the inequalities and approximations that follow.

Let us now study $\log u(s)$ in more detail. Of course, because $\log$ is a monotonic function and $u(s)$ is strictly increasing, $\log u(s)$ must also be strictly increasing, and so $\frac{\deriv}{\deriv s} \log u(s) = u'(s)/u(s)>0$. To better understand the behavior of $\log u(s)$ we can differentiate it directly and simplify to obtain
\begin{equation}
\frac{\deriv}{\deriv s} \log u(s) = \frac{N}{1-e^{Ns}}-\frac{1}{1-e^{s}}.
\end{equation}
From this we immediately see that $\lim_{s\rightarrow-\infty}\frac{\deriv}{\deriv s} \log u(s)=N-1$ and $\lim_{s\rightarrow \infty}\frac{\deriv}{\deriv s} \log u(s)=0$. 

So far we have seen that the limiting values of the derivative of $\log u(s)$ are $N-1$ for large, negative $s$ and $0$ for large, positive $s$. In fact, the derivative of $\log u(s)$ is monotonically decreasing from $N-1$ to $0$, as can be seen by examining the second derivative of $\log u(s)$. In particular, we have
\begin{equation}
\frac{\deriv^{2}}{\deriv s^{2}} \log u(s)
=\frac{N^{2}\,u(s)\,u(-s)-1}{e^{s}+e^{-s}-2}
\end{equation}
which is clearly an even function of $s$, as it must be by Equation~\ref{eq:even}. Furthermore, for $s\neq 0$, $e^{s}+e^{-s}-2=(1-e^s)^{2}/e^{s}>0$ so that the sign of the second derivative depends on the sign of  $N^2\,u(s)\,u(-s)-1$. Now, by Equation~\ref{eq:hyp} in Table~\ref{tab:identities} we have $u(s)u(-s) =\left(\frac{\sinh (s/2)}{\sinh (Ns/2)}\right)^{2}$.  It is easy to show that $\frac{\sinh(s/2)}{\sinh (Ns/2)}$ is a non-negative, even function of $s$ that is decreasing for positive $s$, and has a global maximum at $1/N$ for $s=0$. Thus $N^2\,u(s)\,u(-s)-1$ is always negative for $s\neq 0$ and so is 
$\frac{\deriv^{2}}{\deriv s^{2}} \log u(s)$. Furthermore,  for $s=0$ we have $\frac{\deriv^{2}}{\deriv s^{2}} \log u(s)=-\frac{1}{12}(N^{2}-1)<0$. 

In other words, we have shown that $\frac{\deriv^{2}}{\deriv s^{2}} \log u(s)<0$ for all $s$; we will refer to this condition as the probability of fixation being ``$\log$-concave'' in $s$ and it will serve as the key fact for deriving inequalities in the next section. Although our conclusion that $u(s)$ is $\log$-concave was established using elementary methods, it is worth noting that from a more sophisticated perspective one can see this almost immediately from Identities~\ref{eq:exp} and~\ref{eq:inv}. In particular, Equation~\ref{eq:inv} shows that $1/u(s)$ for $s\geq 0$ is a completely monotone function and hence $\log$-convex for $s\geq0$; it then follows that $u(s)$ is $\log$-concave for $s\geq0$. Furthermore, Equation~\ref{eq:exp} shows that $\log u(s)$ for $s<0$ only differs from $\log u(s)$ for $s>0$ by the addition of a linear function, so that $u(s)$ must be $\log$-concave for $s<0$ as well.

To summarize our results so far, $\log u(s)$ is a monotonically increasing function of $s$ whose derivative decreases monotonically from $N-1$ to $0$ as $s$ goes from $-\infty$ to $\infty$. Furthermore, because the second derivative of $\log u(s)$ is an even function of $s$, this decrease occurs symmetrically around the point $s=0$. This simple behavior allows many immediate insights into the behavior of $\log u(s)$. For instance, we know immediately that the derivative at $s=0$ must be $(N-1)/2$, i.e.~half-way between these limiting values.

\section{Inequalities}
\label{sec:inequalities}

\begin{figure}
\center
\includegraphics[width=6cm]{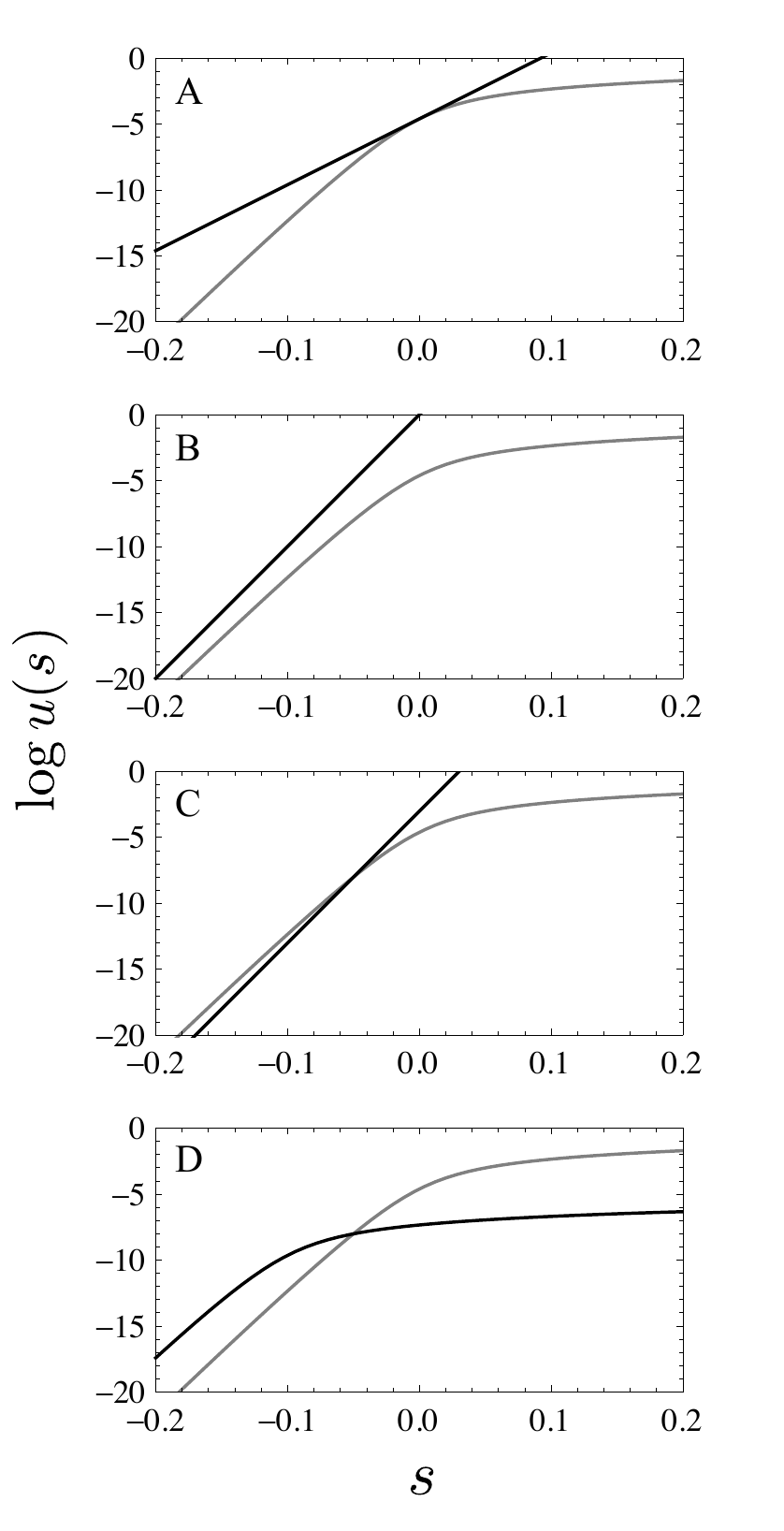}
\caption[]{
A geometric method for developing inequalities for the probability of fixation. For all panels, $N=101$, the gray line is the $\log$ probability of fixation and the black line is another function used to develop an inequality. A) Any tangent of $\log u(s)$ is always greater than or equal to $\log u(s)$. B)~$(N-1)s$ is always greater than the probability of fixation. C) Any line of the form $(N-1)s+b$ intersects $\log u(s)$ at most once and therefore can be used to construct an inequality. D) Any function of the form $u(s+c,N)+b$ intersects $\log u(s)$ at most once and therefore can be used to construct an inequality. 
}
\label{fig:inequalities}
\end{figure}

Our results on the $\log$-concavity of $u(s)$ now allow us to develop a series of inequalities for $u(s)$. The main idea is that the simple
geometry of $\log u(s)$ makes it easy to construct inequalities for $\log u(s)$. Because $\log$ is a monotonically increasing function, these inequalities for $\log u(s)$ also imply inequalities for $u(s)$.

Our first set of inequalities come from the observation that because $\log u(s)$ is concave, any line tangent to the graph of $\log u(s)$ provides an upper bound (see Figure~\ref{fig:inequalities}A). In particular, if we choose the tangent of $\log u(s)$ at $(\tilde{s},\log u(\tilde{s}))$, we have
\begin{equation}
\log u(s) \leq \log u(\tilde{s})+\frac{u'(\tilde{s})}{u(\tilde{s})}(s-\tilde{s}),
\end{equation}
where the inequality is strict for $s\neq \tilde{s}$. Exponentiating then rearranging gives us
\begin{equation}
\label{eq:tan1}
u(s)\leq u(\tilde{s})\,e^{\frac{u'(\tilde{s})}{u(\tilde{s})}(s-\tilde{s})}.
\end{equation}
In the special case $\tilde{s}=0$, this inequality simplifies to
\begin{equation}
u(s)\leq \frac{e^{(N-1)s/2}}{N},
\end{equation}
while letting $\tilde{s}\rightarrow -\infty$ gives us
\begin{equation}
\label{eq:eqineq}
u(s)\leq e^{(N-1)s}
\end{equation}
(see Figure~\ref{fig:inequalities}B).

In fact, this last inequality is strict for all $s$, as can be seen by using Equation~\ref{eq:exp} from Table~\ref{tab:identities} on $e^{(N-1)s}-u(s)$ and putting all the terms over a common denominator. Equation~\ref{eq:eqineq} is particularly useful for understanding the sequential fixations Markov chain because when this Markov chain is reversible its equilibrium distribution is proportional to $\pi_{M}(i) e^{(N-1)\log f(i)}$, where $\pi_{M}(i)$ is the equilibrium frequency of allele $i$ under neutrality and $f(i)$ is the fitness of allele $i$. Since $s$ is a difference between $\log$ fitnesses, we see that Equation~\ref{eq:eqineq} actually describes a relationship between the probability of fixation and the equilibrium distribution of the resulting weak-mutation Markov chain. Elsewhere, we use this inequality and several of the other inequalities below to understand the relationship between the short-term and long-term behavior of the sequential fixations Markov chain under the house of cards model~\citep{McCandlish13e}.

Our second set of inequalities relies on the observation that, because $\log u(s)$ is concave and has a derivative that approaches $N-1$ as $s\rightarrow -\infty$, the derivative of $\log u(s)$ is always less than $N-1$ and thus a line with slope $N-1$ intersects with $\log u(s)$ at most once (Figure~\ref{fig:inequalities}C).

In particular, if a line of the form $(N-1)s+c$ intersects $\log u(s)$ at some point $(\tilde{s},\log u(\tilde{s}))$, then $\log u(s)$ is above $(N-1)s+c$ for $s<\tilde{s}$ and below $(N-1)s+c$ for $s>\tilde{s}$, since the derivative of $(N-1)s+c$ is $N-1$ and hence always greater than the derivative of $\log u(s)$. Solving for $c$ in terms of $\tilde{s}$ gives us $c=\log( u(\tilde{s}))-(N-1)\tilde{s}$. Thus, for $s>\tilde{s}$, we have
\begin{equation}
\log u(s)<(N-1)(s-\tilde{s})+\log u(\tilde{s})
\end{equation}
and the inequality is reversed for $\tilde{s}>s$. Exponentiating this inequality and rearranging yields
\begin{equation}
\label{eq:discrim1}
\frac{u(s)}{u(\tilde{s})}<e^{(N-1)(s-\tilde{s})}
\end{equation}
for $s>\tilde{s}$ and the inequality is reversed for $s<\tilde{s}$. For the special case of $\tilde{s}=0$, this also simplifies to
\begin{equation}
\label{eq:atorigin}
u(s)<\frac{e^{(N-1)s}}{N}
\end{equation}
for $s>0$ and the inequality is reversed for $s<0$.
  
Equation~\ref{eq:discrim1} is interesting because given two possible mutations with selection coefficients $s$ and $\tilde{s}$ that enter the population at the same rate, under weak mutation the odds of the mutation with selection coefficient $s$ next becoming fixed in the population rather than the mutation with selection coefficient $\tilde{s}$ are given by $u(s)/u(\tilde{s})$~\citep{Gillespie83}. Thus, this ratio summarizes the ability of natural selection to discriminate between these two mutations. Equation~\ref{eq:discrim1} then describes a limitation on this ability to discriminate in terms of the population size and the difference between the selection coefficients of the two mutations. 

Given our result on the ratio $u(s)/u(\tilde{s})$, it might also be interesting to understand how this ratio changes if both selection coefficients are increased by some amount $c$, i.e.~$u(s+c)/u(\tilde{s}+c)$. Intuitively, we are asking about how the ability of natural selection to discriminate against two mutations changes if we fix the fitness difference between these mutations but allow their fitnesses to vary relative to the wild-type. 

To construct a relevant inequality, we use the same approach as before, but consider the geometry of $\log u(s)$ and $\log (u(s+c))+d$, that is, we consider the relationship between the curve $\log u(s)$ and a translated version of itself. Because the derivative of both the original curve and its translation are decreasing, the derivative of one of these curves is greater than the other at all $s$ and so the two curves intersect at most once (Figure~\ref{fig:inequalities}D).

In particular, suppose we stipulate that these curves should cross at $(\tilde{s},\log u(\tilde{s}))$. Solving for $d$ yields $d=\log u(\tilde{s}) -\log u(\tilde{s}+c)$. Using this value for $d$, for $c>0$ we have $\frac{\deriv}{\deriv s} \log (u(s+c))+d < \frac{\deriv}{\deriv s} \log u(s)$, since $\frac{\deriv}{\deriv s} \log u(s)$ is decreasing in $s$ and $d$ is negative because $\log u(s)$ is increasing in $s$. Thus, the two curves cross only once and for $c>0$, $\log (u(s+c))+d> \log u(s)$ for $s<\tilde{s}$ and $\log (u(s+c))+d<\log u(s)$ for $s>\tilde{s}$. Exponentiating and rearranging terms, we have:
\begin{equation}
\label{eq:discrim2}
\frac{u(s)}{u(\tilde{s})}>\frac{u(s+c)}{u(\tilde{s}+c)}
\end{equation}
for $c>0$ and $s>\tilde{s}$; for $s<\tilde{s}$ the inequality in Equation~\ref{eq:discrim2} is reversed. Of course, one could construct an analogous equation for $c<0$ (which, incidentally, could also be used to derive Equation~\ref{eq:discrim1} by taking the limit as $c\rightarrow -\infty$).

What Equation~\ref{eq:discrim2} tells us is that increasing the selection coefficients of two mutations by equal amounts always decreases the ability of natural selection to discriminate between them. To put this another way, in the regime where new mutations are entering the population one at a time, natural selection's power to discriminate between these mutations based on their fitnesses is maximized for strongly deleterious mutations. 

All of our results so far have been presented for the probability of fixation of a new mutation because the primary intended application of these results has been studying evolution under weak mutation. However, for some applications, it is interesting to also consider the probability of fixation of an allele that begins at a frequency greater than $1/N$. The results for this more general case are almost completely analogous with those given above; see~\ref{sec:polymorphic}.

\section{Approximations}

Our strategy for developing inequalities was based in the observation that the $\log$ probability of fixation has many useful properties that stem from the basic identity $u(s)/u(-s)=e^{(N-1)s}$. Besides conferring certain formal properties on the logarithm of the probability of fixation, this identity also suggests a natural strategy for developing approximations to the probability of fixation. In particular, by rearranging this identity to read $u(-s)=u(s)e^{-(N-1)s}$, we can take an approximation defined for positive selection coefficients and use it to derive a corresponding approximation for negative selection coefficients. Furthermore, an
approximation constructed in this manner automatically preserves the reversibility and equilibrium distribution of any weak-mutation Markov chain
based on such an approximation. The same idea can be applied if one has a good approximation for negative selection coefficients. Our main goal in this section is to use this strategy to develop a number of useful approximations for $u(s)$, while comparing the performance of these approximations to other, previously proposed, approximations. 

Historically, most work on approximate formulas for the probability of fixation has concerned the probability of fixation under the Wright-Fisher process or its diffusion limit. Because there is no known exact formula for the probability of fixation for a finite population evolving under a Wright-Fisher process, analytically assessing the accuracy of such approximations is quite difficult~\citep[although see][see also~\citealt{Gale90} for a more accessible treatment]{Moran60,Burger95}. Having an exact formula against which to measure the accuracy of approximations thus provides one of our major motivations for working in a Moran process framework. Fortunately, because the exact formula for the probability of fixation of a Moran process coincides up to a rescaling of variables with the formula for the diffusion limit of a Wright-Fisher process~\citep{Kimura57,Kimura62}, it is easy to adapt previously proposed approximations to the current context. In what follows we will therefore present these adapted expressions while providing citations for the original formulae.

The most commonly used approximation for the probability of fixation of a new mutation was given by~\citet{Fisher30} and~\cite{Wright31} as:
\begin{equation}
w_{N}(s)=\frac{s}{1-e^{-Ns}}.
\end{equation}
Let us begin by considering the behavior of this expression in light of our previous analysis of $u(s)$. Suppressing the $N$ in our notation as for $u(s)$, we see that $w(s)$ can be derived from the expression for $u(s)$ using the approximation $1-e^{-s}\approx s$, valid for small $|s|$, so that we expect $w(s)$ to be a good approximation for small $|s|$, but not necessarily for large $|s|$. Indeed, for large, positive $s$, $w(s)$ can be greater than 1 and therefore not a valid probability. More precisely:
\begin{equation}
\label{eq:werror}
\frac{w_{N}(s)}{u_{N}(s)}=\frac{s}{1-e^{-s}}
\end{equation}
so that the error in using $w(s)$ is solely a function of $s$ (and not $N$) and $w(s)$ overestimates the probability of fixation for beneficial mutations and underestimates the probability of fixation for deleterious mutations. Since $w(s)$ becomes more accurate as $|s|$  becomes small, if we fix $Ns=S$ and let $N\rightarrow \infty$, we have the well-known result:
\begin{equation}
\label{eq:wS}
\lim_{N\rightarrow \infty} N{u_{N}(s)}=\lim_{N\rightarrow \infty} N{w_{N}(s)}=\frac{S}{1-e^{-S}}.
\end{equation}
Because it is often useful to have an approximation for $u_{N}$ such that the corresponding approximation for $Nu_{N}(s)$ can be expressed solely in terms of the compound parameter $S=Ns$ as $N\rightarrow \infty$, we will highlight this characteristic when it occurs in the approximations that follow.
Table~\ref{tab:widentities} gives several useful identities for $w(s)$.

\setid

\begin{table}[p]
\begin{align}
\label{eq:kexp} &\frac{w(s)}{w(-s)} =e^{Ns}  & &   \text{Reversibility} \\
\label{eq:series} & w(s)=\left\{ \begin{aligned}  s\sum_{k=0}^{\infty}e^{-k\,Ns} && \text{for } s>0 \\ -s\sum_{k=1}^{\infty}e^{k\,Ns} && \text{for } s<0 \end{aligned} \right. & & \text{Geometric Series}\\
\label{eq:kcomp} & w(s)-s=w(-s)  & & \text{Comparison with $Ns\gg1$ limit} \\
\label{eq:kderiv} & w'(s)=1-w'(-s)  & & \text{First derivative} \\
\label{eq:kderiv2} & w''(s)=w''(-s)  & & \text{Second derivative} \\
\label{eq:khyp1} & w(s)=e^{Ns/2}\left(\frac{s/2}{\sinh(Ns/2)}\right) & & \text{Hyperbolic identity} \\
\label{eq:khyp2} & w(s)w(-s)=\left(\frac{s/2}{\sinh(Ns/2)}\right)^{2} & & \text{Hyperbolic identity II} \\
\label{eq:khyp3} & w(s)+w(-s)=s\coth(Ns/2) & & \text{Hyperbolic identity III} \\
\label{eq:khyp4} & w(s)=\frac{s}{2}\left(1+\coth(Ns/2)\right) & & \text{Hyperbolic identity IV}
\end{align}
\caption{\label{tab:widentities}Basic identities for \citet{Fisher30} and~\citet{Wright31}'s approximation to the probability of fixation} 
\end{table}
\afterpage{\clearpage}

\setreg

While $w(s)$ shares many of the properties of $u(s)$ with regard to its logarithm, it is also useful because it is significantly easier to analyze directly. Turning first to $\log w(s)$, all of the relations in Equations~\ref{eq:linerar},~\ref{eq:derivlog} and~\ref{eq:even} continue to hold, but with $\nu(N)=N$ instead of $N-1$. Furthermore, we have
\begin{gather}
\frac{\deriv}{\deriv s} \log w(s) = \frac{N}{1-e^{Ns}}+\frac{1}{s},\\
\frac{\deriv}{\deriv s^{2}} \log w(s) = \frac{N^{2}\,w(s)\,w(-s)-1}{s^{2}}
\end{gather}
from which one can show (using Equation~\ref{eq:khyp2} from Table~\ref{tab:widentities}) that $w(s)$ is $\log$-concave, that the derivative of $\log w(s)$ decreases monotonically from $N-1$ to $1$ as $s$ increases from $-\infty$ to $\infty$, and that this decrease in the derivative occurs symmetrically around the point $s=0$. The $\log$ concavity of $w(s)$ also makes it easy to construct inequalities for $w(s)$, as in the previous section.

\begin{figure}
\center
\includegraphics[width=8cm]{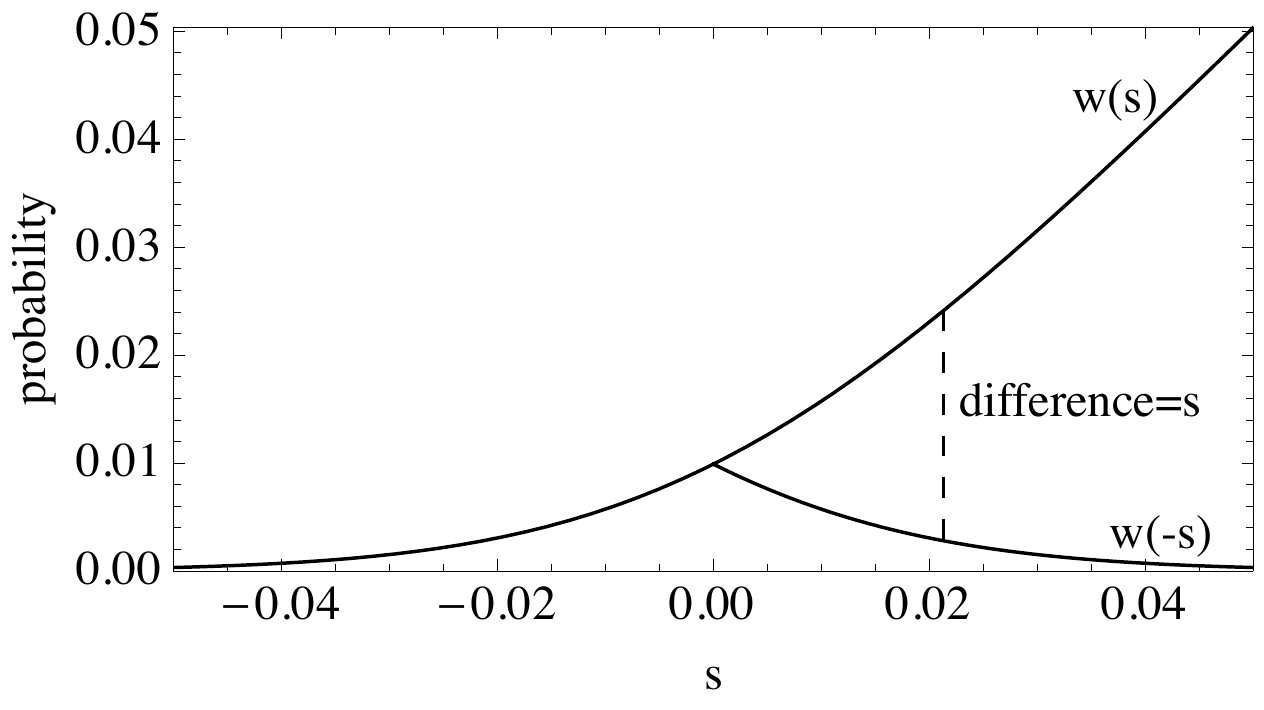}
\caption[]{\label{fig:ID1}
Graphical representation of the identity given by Equation~\ref{eq:kcomp} for $N=101$. The difference in the probability of fixation between a mutation with selection coefficient $s$ and selection coefficient $-s$ is equal to $s$, which for $s>0$ is also equal to the probability of fixation given by $w$ for an infinite population (i.e.~the large $Ns$ limit).
}
\end{figure}

The direct analysis of $w(s)$ is much simpler than the direct analysis of $u(s)$ as a formal consequence of the remarkable identity given by Equation~\ref{eq:kcomp} in Table~\ref{tab:widentities}. But before we explore the formal implications of~Equation~\ref{eq:kcomp}, it is important to understand what this identity means biologically. In particular,~Equation~\ref{eq:kcomp} expresses the probability of fixation given by $w_{N}(s)$ for positive $s$ as the sum of the probability of fixation for that selection coefficient in an infinite population ($\lim_{N\rightarrow \infty}w_{N}(s)=s$) plus some extra probability due to the effects of finite population size. The reason this identity is remarkable is not because such a decomposition is possible, but because the amount of extra probability is symmetrical around $s=0$: the extra probability due to finite population size for $s>0$ is precisely equal to the probability of fixation of a deleterious allele whose selection coefficient has the same absolute value (Figure~\ref{fig:ID1}). Note that this relationship holds approximately for $u(s)$ as well (Equation~\ref{eq:almostsym} from Table~\ref{tab:identities}).

Formally,~Equation~\ref{eq:kcomp} means that the derivatives of $w(s)$ must respect certain symmetries (Equations~\ref{eq:kderiv} and~\ref{eq:kderiv2}). The first and second derivatives are given by
\begin{gather}
 w'(s) = \frac{w(s)}{s} \left(1- N\,w(-s)\right), \\
 w''(s)  = \left(\frac{N}{s}\right)^{2}\left( w(s)\,w(-s)\right)\left(w(s)+w(-s)-2/N\right)
\end{gather}
from which it is easy to show that $w(s)$ is convex and increasing in $s$ (to show convexity, note that $w(s)+w(-s)=s \coth(N s/2)$, which clearly has a minimum at $s=0$; the value at this minimum is $2/N$, as can be shown by expressing $\coth$ in terms of exponentials and then using L'H\^{o}pital's rule). In particular, $w'(s)$ increases from $0$ to $1$ as $s$ increases from $-\infty$ to $\infty$, and, by Equation~\ref{eq:kderiv2}, this change occurs in a manner that is symmetric around $s=0$. The convexity of $w(s)$ likewise enables the construction of inequalities for $w(s)$ as in the previous section, but where the inequalities are constructed by considering the behavior of $w(s)$ directly rather than its $\log$.

As we have seen, $w(s)$ has many useful properties. However, further approximations are often necessary because of the $1-e^{-Ns}$ term in the denominator, which frequently results in expressions that are difficult to work with. One common simplification of $w(s)$ is to assume that $Ns$ is large so that $1-e^{-Ns}\approx 1$ and therefore $u(s)\approx s$ for $s>0$ and $0$ otherwise. While $s$ is certainly easy to integrate against and shares the appealing property of $w(s)$ that it can often be used to write the evolutionary dynamics in terms of $S=Ns$, the fact that deleterious fixations are treated as impossible under this approximation sometimes produces pathological consequences~\citep[for instance, the prediction that fitness should increase indefinitely over time instead of reaching mutation-selection-drift balance; see][for a discussion of other related issues]{McCandlish13e}. Thus, it may be useful to extend this approximation to include the possibility of deleterious fixations.

The most natural way to conduct such an extension is to stipulate that the property $u(s)/u(-s)=e^{(N-1)s}$ should be maintained, which also leaves the equilibrium distribution of the sequential fixations Markov chain unchanged. Thus, if $u(s)\approx s$ for $s>0$, then we must have $u(-s)\approx s e^{-(N-1)s}$ for $s>0$ or equivalently:
\begin{equation}
\label{eq:currentapprox}
u(s)\approx \left\{\begin{aligned} & s && \text{for } s>0 \\
&-s\,e^{(N-1)s} && \text{for } s \leq 0.
\end{aligned} \right. 
\end{equation}
This approximation is particularly useful because the term $e^{(N-1)s}$ in the probability of fixation for a deleterious mutant means that the integrals produced using this approximation can be expressed in terms of the equilibrium distribution of the sequential fixations Markov chain~\citep[see, e.g., the supplemental material to][]{McCandlish13e}. For fixed $S=Ns$ and large $N$, we also have:
\begin{equation}
Nu_{N}(s) \approx \left\{\begin{aligned} & S && \text{for } S>0 \\
&-S\,e^{S} && \text{for } S \leq 0.
\end{aligned} \right.
\end{equation}

One defect of Equation~\ref{eq:currentapprox} is that for large $s$, the resulting expression can be greater than 1. This can be rectified by considering the true large $N$ limit of $u_{N}(s)$, which is is $1-e^{-s}$ for $s>0$ and $0$ otherwise. Extending this approximation to negative $s$ gives us:
\begin{equation}
\label{eq:currentapproxorr}
u(s)\approx \left\{\begin{aligned} & 1-e^{-s} && \text{for } s>0 \\
&e^{(N-1)s}-e^{Ns} && \text{for } s \leq 0.
\end{aligned} \right. 
\end{equation}

A defect shared by these approximations based on the large $N$ limit of $u_{N}(s)$ is that they grossly underestimate the probability of fixation of nearly neutral mutations (e.g.~they approximate the probability of fixation for a strictly neutral mutation as $0$ instead of $1/N$). \citet{Poon00} proposed a solution to this problem based on the strategy of including an additive correction term that (1) is some multiple of an exponential function (and therefore easy to integrate against), (2) results in the approximation having the correct value for $s=0$ and, (3) results in the approximation having the correct derivative with respect to $s$ when evaluated at $s=0$. For the approximation $u(s)\approx s$ for $s>0$ and $0$ otherwise, this strategy results in the approximation~\citep{Poon00}:  
\begin{equation}
\label{eq:poonotto}
u(s)\approx \left\{\begin{aligned} & s+\frac{1}{N}e^{-(N+1)s/2} && \text{for } s > 0\\
 &\frac{1}{N}e^{(N-1)s/2} && \text{for } s\leq0,
\end{aligned} \right. 
\end{equation}
or for fixed $S$ and large $N$:
\begin{equation}
\label{eq:poonottoS}
Nu_{N}(s)\approx \left\{\begin{aligned} & S+e^{-S/2} && \text{for } S > 0\\
 &e^{S/2} && \text{for } S\leq0.
\end{aligned} \right. 
\end{equation}

While the~\citet{Poon00} approximation provides a very nice approximation for advantageous mutations of small effect, the approximation for deleterious mutations is rather poor and the use of this approximation does not maintain the equilibrium distribution of the weak-mutation dynamics achieved under the exact expression. We can fix both defects using our strategy for developing approximations based on the relationship $u(s)/u(-s)=e^{(N-1)s}$. Using the~\citet{Poon00} approximation for positive $s$ to derive the corresponding approximation for negative $s$ gives us:
\begin{equation}
\label{eq:currentpoonotto}
u(s)\approx \left\{\begin{aligned} & s+ \frac{1}{N}e^{-(N+1)s/2} && \text{for } s > 0\\
  &-s\,e^{(N-1)s}+\frac{1}{N}e^{(3N-1)s/2} && \text{for } s\leq0.
\end{aligned} \right. 
\end{equation}
For fixed $S=Ns$ and large $N$, this simplifies to
\begin{equation}
\label{eq:currentpoonottoS}
Nu_{N}(s)\approx \left\{\begin{aligned} & S+ e^{-S/2} && \text{for } S > 0\\
  &-S\,e^{S}+e^{3S/2} && \text{for } S\leq0.
\end{aligned} \right. 
\end{equation}
We can also modify this approximation to give better results for large $s$ in the manner of Equation~\ref{eq:currentapproxorr}:
\begin{equation}
\label{eq:currentpoonottoorr}
u(s)\approx \left\{\begin{aligned} & 1-e^{-s}+ \frac{1}{N}e^{-(N+1)s/2} && \text{for } s > 0\\
  & e^{(N-1)s}-e^{Ns} +\frac{1}{N}e^{(3N-1)s/2} && \text{for } s\leq0. 
\end{aligned} \right.
\end{equation}

An additional strength of several of the above approximations is that the approximations provide bounds on the true probability of fixation. In particular, Equation~\ref{eq:currentapproxorr} provides a lower bound on $u(s)$ while Equations~\ref{eq:poonotto},~\ref{eq:currentpoonotto}, and~\ref{eq:currentpoonottoorr} provide upper bounds. To see why this is true, we first note that for $s>0$, we have:
\begin{gather}
1-e^{-s}< s < w(s)\label{eq:ineqs1} \\
1-e^{-s}< u(s) < w(s) \\
1-e^{-s} < u(s) < 1-e^{-s}+\frac{e^{-(N+1)s/2}}{N}<s+\frac{e^{-(N+1)s/2}}{N} \label{eq:ineqs3}\\
1-e^{-s}< s<s+\frac{e^{-(N+1)s/2}}{N}.
\end{gather}
These inequalities are mostly obvious, but two require additional comment. First, $u(s)<w(s)$ for $s>0$ follows from Equation~\ref{eq:werror}. Second, it is not immediately clear that $ u(s) < 1-e^{-s}+\frac{e^{-(N+1)s/2}}{N}$ for $s>0$. To see that this is true, first note that using Equation~\ref{eq:hyp}:
\begin{equation}
\left(1-e^{-s}+\frac{e^{-(N+1)s/2}}{N} \right)/u(s) =1+e^{-Ns}\left(\frac{\sinh(Ns/2)-N\sinh(s/2)}{N \sinh(s/2) }\right).
\end{equation}
Clearly whether this ratio is greater or less than one only depends on the sign of the numerator of the fraction in parentheses on the right hand side, since the denominator is always positive for $s>0$. Noting that $\sinh(0)=0$ and $\frac{d}{dx} \sinh(x)=\cosh(x)$, where $\cosh(x)$ is non-negative and increasing for $x>0$, we conclude that $\sinh(cx)>c\sinh(x)$ for all $c,x>0$, so that the numerator is positive as required.

Now, it still remains to show that Equation~\ref{eq:currentapproxorr} provides a lower bound on $u(s)$ while Equations~\ref{eq:poonotto},~\ref{eq:currentpoonotto}, and~\ref{eq:currentpoonottoorr} provide upper bounds. The fact that Equation~\ref{eq:poonotto} is an upper bound follows from the inequality in Equation~\ref{eq:ineqs3} for the case $s>0$ and by Equation~\ref{eq:atorigin} for the case $s\leq 0$. For Equations~\ref{eq:currentapproxorr},~\ref{eq:currentpoonotto}, and~\ref{eq:currentpoonottoorr} the direction of the bound for $s>0$ follows by the inequalities in Equations~\ref{eq:ineqs1} and~\ref{eq:ineqs3} for $s>0$ and can easily be checked for $s=0$. For the case $s<0$ we simply note that because the values of these approximations for $s<0$ were determined from the values for $s>0$ and the relation $u(s)/u(-s)=e^{N-1(s)}$, the proportional error in the approximation is symmetric around $s=0$, so that the error for $s<0$ must be in the same direction as the error for $s>0$. These bounds are all strict except that Equations~\ref{eq:poonotto},~\ref{eq:currentpoonotto}, and~\ref{eq:currentpoonottoorr} are equal to $u(s)$ at $s=0$.

Finally, while the above inequalities cannot be used to establish a relationship between the approximation in Equation~\ref{eq:currentpoonotto} and $w(s)$, Equation~\ref{eq:currentpoonottoS} does provide an upper bound for $Nw_{N}(s)$ in the large $N$, fixed $Ns$ limit (Equation~\ref{eq:wS}), in the sense that
\begin{equation}
\label{eq:currentpoonottoineq}
\frac{S}{1-e^{-S}}\leq \left\{\begin{aligned} & S+e^{-S/2} && \text{for } S > 0\\
  &-S\,e^{S}+e^{3S/2} && \text{for } S\leq0.
\end{aligned} \right. 
\end{equation}
To see that the first inequality is true for $S>0$, note that
\begin{align}
\left(S+e^{-S/2}\right)/\frac{S}{1-e^{-S}} &=1-e^{-S}-\frac{e^{-3 S/2}}{S}+\frac{e^{-S/2}}{S}\\
 &=1+\frac{e^{-3S/2}}{S}\left(-1 + e^S - Se^{S/2}\right)
\end{align}
Thus, it suffices to show that $-1 + e^S - Se^{S/2}\geq 0$ for $S>0$. Now, using the series expansion for the exponential function, we have:
\begin{align}
-1 + e^S - Se^{S/2} &=\left(\sum_{k=1}^{\infty}S^{k}/k!\right)-S\left(\sum_{k=0}^{\infty}(S/2)^{k}/k! \right) \\
 &= \sum_{k=1}^{\infty}S^{k}\left(\frac{1}{k!} -\frac{(1/2)^{k-1}}{(k-1)!}\right) \\
 &=  \sum_{k=1}^{\infty}S^{k}\left(\frac{1-k (1/2)^{k-1}}{k!} \right) \\
 &> 0
\end{align}
where the last line follows because the terms in the sum are zero for $k=1,2$ and positive for $k\geq3$ since $2^{k-1}>k$ for $k\geq 3$.
The inequality for $S<0$ then follows because
\begin{equation}
\left(S+e^{-S/2}\right)/\left(\frac{S}{1-e^{-S}} \right)=\left(-(-S)\,e^{(-S)}+e^{3(-S)/2}\right)/\left(\frac{(-S)}{1-e^{-(-S)}}\right) \quad \text{for } S>0,
\end{equation}
so that the proportional error is symmetric around $S=0$. Finally, using L'H\^{o}pital's rule, one can show that the left hand side of Equation~\ref{eq:currentpoonottoineq} is equal to 1 for $S=0$ (as is the right hand side), so that not only does the inequality in Equation~\ref{eq:currentpoonottoineq} hold for all $S$, but it is strict for $S \neq 0$.

A different, but still commonly used, strategy for constructing approximations to the probability of fixation is based on using the tangent of $u(s)$ at $s=0$ as an approximation of $u(s)$ for small $s$:
\begin{equation}
u(s)\approx \frac{1}{N} +\left(\frac{N-1}{N}\right) \frac{s}{2},
\end{equation}
or for fixed $Ns$ and large $N$:
\begin{equation}
Nu_{N}(s) \approx 1+\frac{s}{2},
\end{equation}
based on the more general expression given by~\citet{Robertson60}.~\citet{Hill82} later modified this approximation to provide more accurate results for advantageous mutations:
\begin{equation}
u(s)\approx \left\{\begin{aligned}  & s && \text{for } s \geq 2\frac{1}{N+1}\\
& \frac{1}{N}+\frac{N-1}{N} \frac{s}{2} && \text{for } -2\frac{1}{N-1}<s < 2\frac{1}{N+1}\\ 
&0 && \text{for } s \leq -2\frac{1}{N-1},\\ \end{aligned} \right.
\end{equation}
or for large $N$ and fixed $S=Ns$
\begin{equation}
Nu_{N}(s) \approx \left\{\begin{aligned}  & S && \text{for } S \geq 2\\
& 1+\frac{S}{2} && \text{for } -2<S < 2\\ 
&0 && \text{for } S \leq -2.\\ \end{aligned} \right.
\end{equation}

The approach towards constructing inequalities in the previous section suggests an obvious parallel to this strategy: by using piecewise linear approximations to $\log u(s)$, one can construct a piecewise exponential approximation to $u(s)$. For instance, the following approximation both preserves the relation $u(s)/u(-s)=e^{(N-1)s}$ and serves as an upper bound on the probability of fixation:
\begin{equation}
u(s)\approx \left\{\begin{aligned}  
& 1 && \text{for } s \geq 2\frac{\log N}{N-1} \\
& \frac{e^{(N-1)s/2}}{N} && \text{for }  -2\frac{\log N}{N-1}<s < 2\frac{\log N}{N-1}\\ 
&e^{(N-1)s} && \text{for } s \leq -2\frac{\log N}{N-1}.\\
 \end{aligned} \right.
\end{equation}
For large $N$ and fixed $Ns=S$, this approximation simplified to
\begin{equation}
\label{eq:expsmall}
N u_{N}(s) \approx e^{S/2},
\end{equation}
which is also an upper bound for Equation~\ref{eq:wS}. Equation~\ref{eq:expsmall} has been used previously in the literature~\citep[e.g., the supplemental material to][]{Chevin14}; see also~\citet{Knudsen05} who relate Equation~\ref{eq:expsmall} to ``generalized weighted frequencies (+gwF)'' models in molecular evolution~\citep{Goldman02}.

Finally, our analysis of the derivatives of $w(s)$ suggests a completely different approach towards constructing approximations: because the slope of $w(s)$ changes monotonically from $0$ to $1$ as $s$ goes from $-\infty$ to $\infty$, and this change is symmetric around $s=0$, we can approximate $w''(s)$ as the cumulative distribution function of a probability distribution, and then integrate to produce an approximation for $u(s)$. For instance, approximating $w''_{N}(s)$ by matching the first two moments suggest using a normal distribution with mean 0 and standard deviation $\sqrt{2/3}\,\pi/N$ or a logistic distribution with mean 0 and scale equal to $\sqrt{2}/N$. For the normal distribution, this yields the approximations:
\begin{equation}
u(s) \approx \frac{2}{3}\pi^{2}\phi(Ns)/N+s\,\Phi(Ns)
\end{equation}
and for fixed $Ns=S$
\begin{equation}
N u_{N}(s) \approx  \frac{2}{3}\pi^{2}\phi(S)+S\,\Phi(S)
\end{equation}
where $\phi$ is the PDF and $\Phi$ is the CDF of a normal distribution with mean 0 and standard deviation $\sqrt{2/3}\pi$. For the logistic distribution, the corresponding approximations are:
\begin{equation}
u(s) \approx \frac{\sqrt{2}}{N} \log\left(1+e^{\frac{Ns}{\sqrt{2}}}\right)
\end{equation}
and for fixed $Ns=S$
\begin{equation}
N u_{N}(s) \approx \sqrt{2} \log\left(1+e^{\frac{S}{\sqrt{2}}}\right).
\end{equation}

\begin{figure}
\hspace{-1cm}\includegraphics[width=7in]{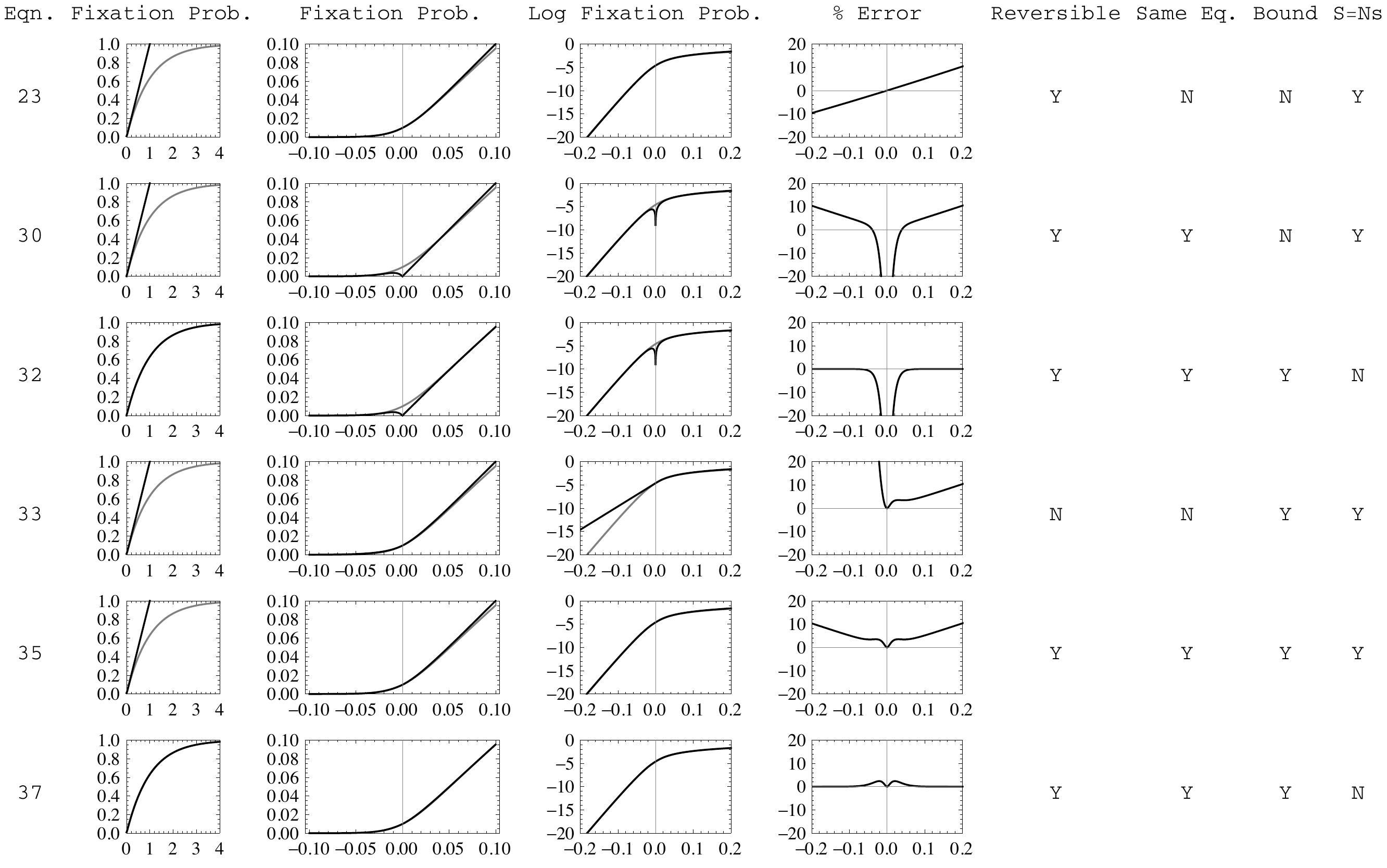}
\caption[]{\label{fig:approx1}
Approximations and their characteristics. For each equation, the first three graphs show the approximation (black curve) and the true probability of fixation (gray curve) as a function of $s$ (x-axis) for $N=101$; the first graph shows the behavior for large $s$, the second graph shows the behavior for small $|s|$, and the third graph shows the $\log$ probability of fixation. The fourth graph shows the percent error of the approximation as a function of $s$. The columns at the right describe whether the approximation has various formal properties (``Y'' for yes; ``N'' for no). The first of these columns asks if the probability of fixation produces a reversible Markov chain when used to construct a sequential fixation Markov chain (i.e.~does it satisfy Equation~\ref{eq:reversableQ} for some choice of $\nu(N)$?). The second column asks if the equilibrium distribution of the sequential fixations Markov chain is the same as under the exact expression (i.e.~does $\nu(N)=N-1$ in Equation~\ref{eq:reversableQ}?). The third column asks if the approximation provides either an upper or lower bound for the probability of fixation. The last column asks if there is a form of the approximation such that the corresponding approximation of $N u_{N}(s)$ can be expressed solely in terms of the compound parameter $S=Ns$ when $N$ is large.
}

\end{figure}

\begin{figure}
\hspace{-1cm}\includegraphics[width=7in]{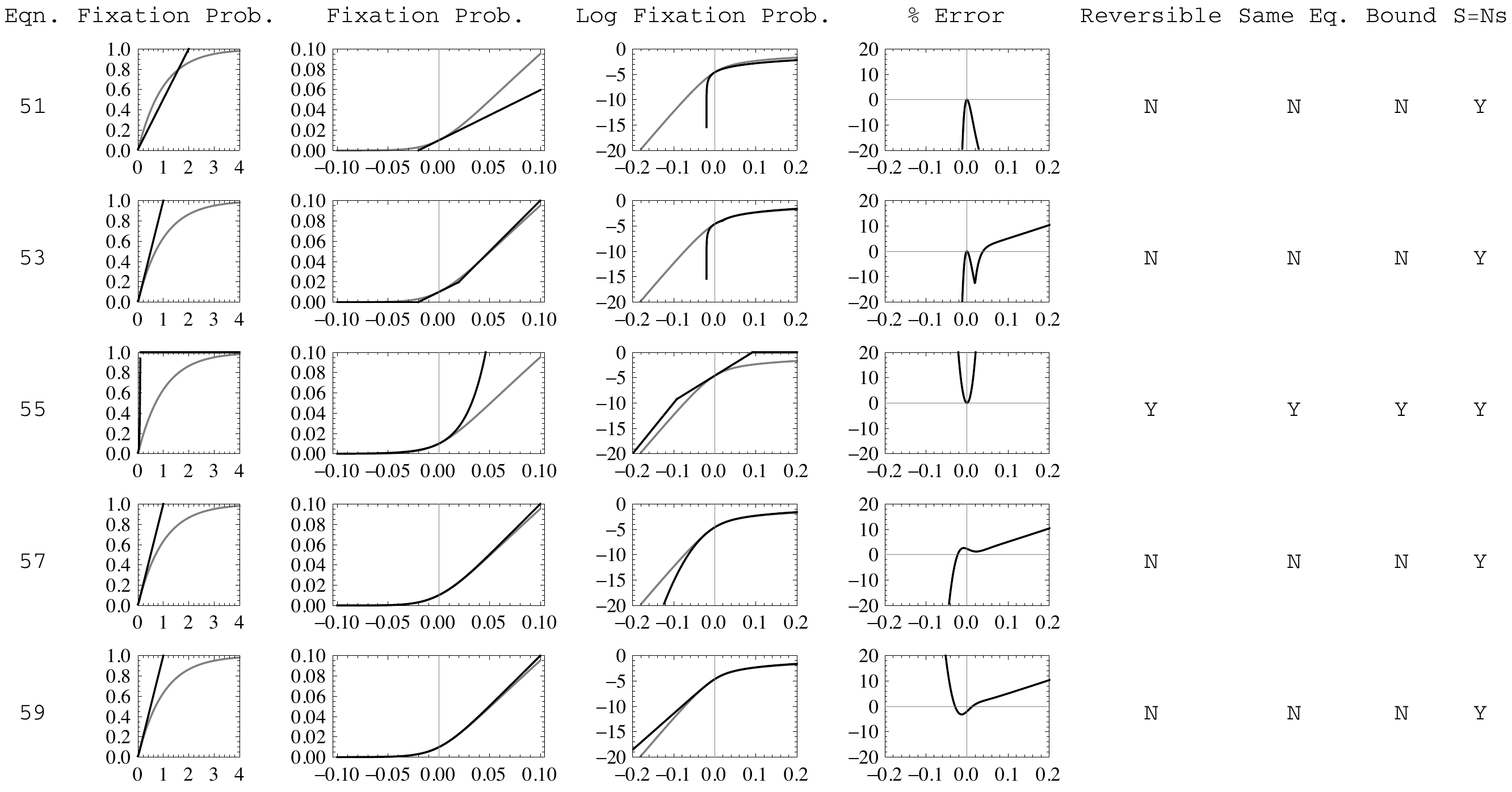}
\caption[]{\label{fig:approx2}
Approximations and their characteristics. The interpretation is the same as for Figure~\ref{fig:approx1}. Note that the approximations in the first two rows both become negative for sufficiently negative $s$; thus, the $\log$ of the probability of fixation under the approximation reaches $-\infty$ at a finite value of $s$, which explains the vertical lines in the third graph.
}
\end{figure}

Figures~\ref{fig:approx1} and~\ref{fig:approx2} provide a summary of the various approximations discussed in this section. While many of these approximations are likely to be useful in some circumstances, the approximation given by Equation~\ref{eq:currentpoonotto} has an accuracy comparable to that of $w(s)$ as well as many useful formal features. In particular, Equation~\ref{eq:currentpoonotto} has a convenient functional form, preserves the structure of the equilibrium distribution of the sequential fixations Markov chain, provides an upper bound on the probability of fixation and can be recast in terms of $S=Ns$ when $S$ is large. It therefore seems appropriate to consider the accuracy of this approximation in somewhat more detail.

\begin{figure}
\center
\includegraphics[width=8cm]{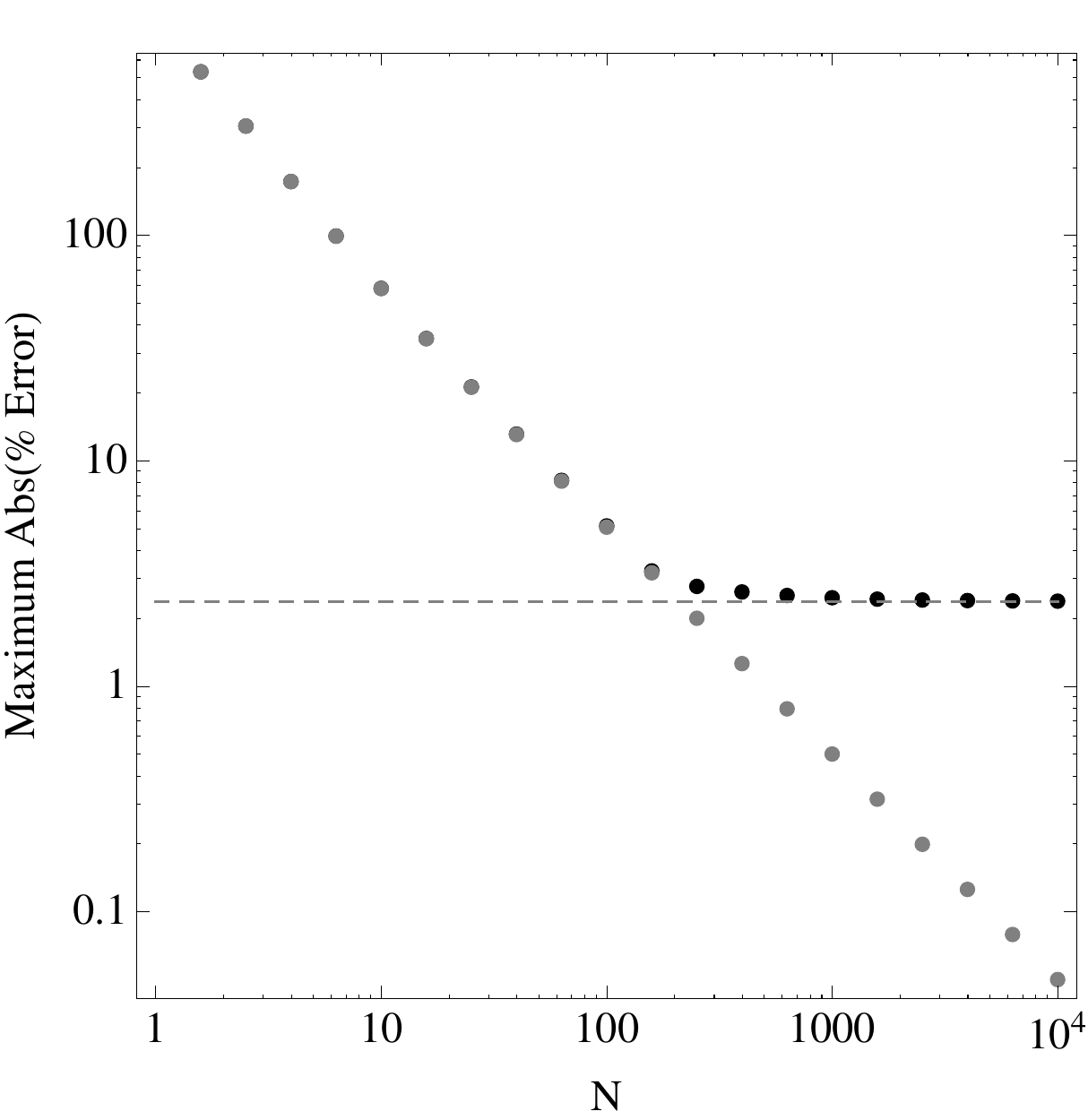}
\caption[]{\label{fig:comp1}
Maximum of the absolute percent error for $|Ns|\leq10$ of $w(s)$ (gray) and Equation~\ref{eq:currentpoonotto} (black) as a function of $N$. The accuracies of the two approximations are very similar for small $N$ but diverge for large $N$. The dashed line shows the large $N$ asymptotic value of the error for Equation~\ref{eq:currentpoonotto}.
}
\end{figure}

\begin{figure}
\center
\includegraphics[width=8cm]{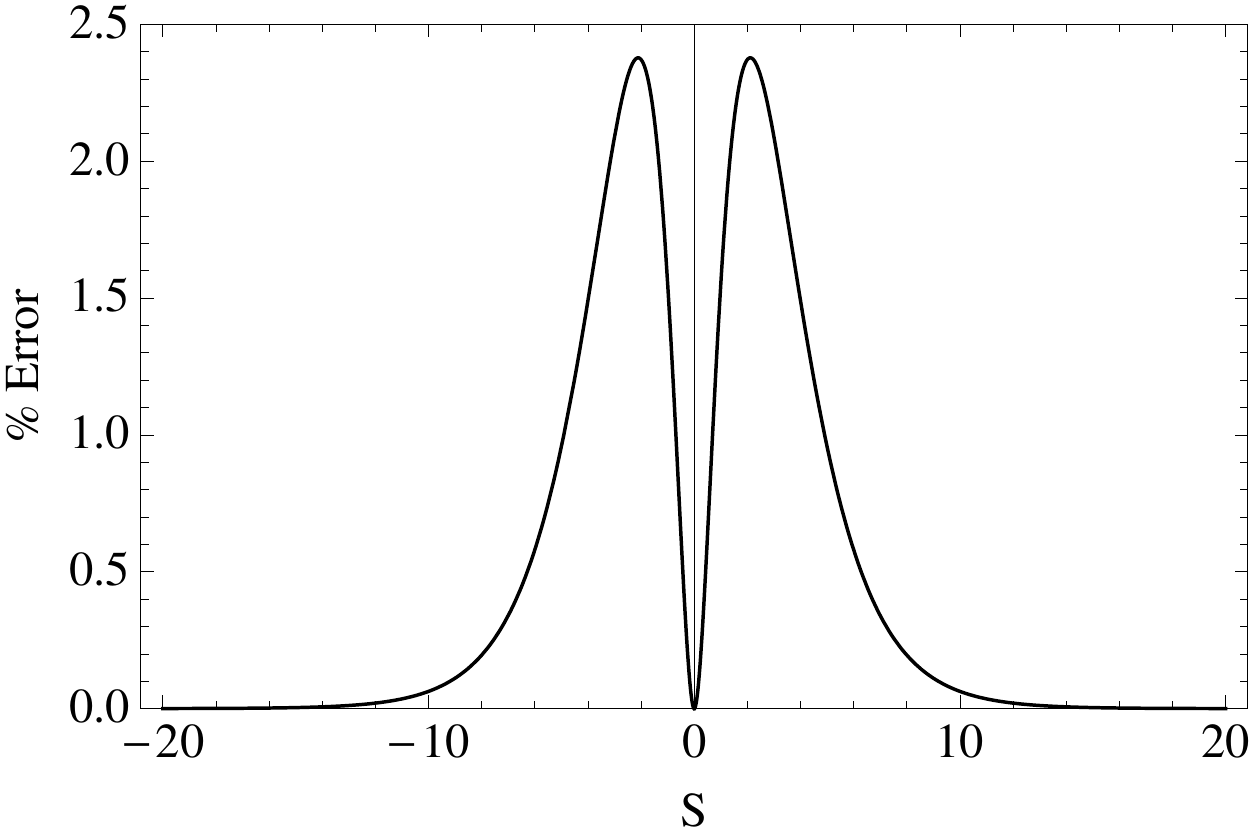}
\caption[]{\label{fig:comp2}
Percent error of Equation~\ref{eq:currentpoonottoS} (black) as a function of $S=Ns$. The maximal error of $\approx 2.38\%$ occurs at $|Ns|\approx2.1$.
}
\end{figure}

Figure~\ref{fig:comp1} compares the accuracy of $w(s)$ and Equation~\ref{eq:currentpoonotto} as a function of $N$ in terms of the maximum absolute value of the percent error for any $s$ in the range $ |Ns| \leq 10$. While the accuracy of $w(s)$ gets better and better for large $N$ (as it must by Equation~\ref{eq:wS}), the accuracy of Equation~\ref{eq:currentpoonotto} decreases initially, but then asymptotes at $\approx 2.38\%$. To understand the source of this error, it is helpful to look at the large $N$, fixed $Ns$ limit, and compare Equation~\ref{eq:currentpoonottoS} (i.e.~the large $N$, fixed $Ns$ version of Equation~\ref{eq:currentpoonotto}) with the exact value given by Equation~\ref{eq:wS}. Figure~\ref{fig:comp2} shows the error in using this approximation as a function of $Ns$. The figure shows that Equation~\ref{eq:currentpoonottoS} is extremely accurate for mutations that are strongly beneficial or deleterious, as well mutations that are very close to neutral, but overestimates the probability of fixation of slightly advantageous and slightly disadvantageous mutations, with a peak error of $\approx 2.38\%$ at $|Ns|\approx2.1$. Because this magnitude of error is acceptable in most circumstances, Equation~\ref{eq:currentpoonotto} provides a very reasonable alternative to $w(s)$ when approximating the probability of fixation.

\section{Applications}
\label{sec:applications}

In order to demonstrate the utility of the results presented so far, we will apply them towards a common goal in molecular evolution: determining the rate of evolution. In the theoretical literature, this rate is typically calculated based on the assumptions that mutations are entering the population with selection coefficients drawn from some distribution and that each such new mutation is lost or fixed independently from the others.  Using the approximation $N u_{w}(s) \approx N w_{N}(s)$ for large $N$ and $S=Ns$ fixed, we will write $W(S)=Nw_{N}(s)$. Under this approximation, the rate of evolution is given by
\begin{equation}
\label{eq:Ohta}
K(X)=\int_{-\infty}^{\infty} W(Y-X)\psi(Y)\,dY
\end{equation}
\citep{Ohta77,Kimura79} where $\psi(Y)\,dY$ is the probability of a mutation occurring with scaled Malthusian fitness in the interval $[Y,Y+dY]$, $X$ is the current scaled fitness of the population and time is measured in units of the inverse of the mutation rate. In other words, $K(X)$ is the instantaneous substitution rate for a population with fitness $X$ when the distribution of fitnesses introduced by mutation has a probability density function given by $\psi$ and time is measured in the expected number of substitutions that would have occurred if all mutations were neutral. 

Our results on the shape of the probability of fixation can provide some immediate insights into this formula. For instance, one natural question is how the substitution rate changes as the current fitness, $X$, changes. It is easy to show that $K(X)$ is decreasing in $X$~\citep[see, e.g.][]{McCandlish13e}, but can we say more than that? Now, suppose that $\psi$ is itself $\log$-concave, as is true for many commonly used distributions (e.g.~normal, exponential, uniform on the interval $[a,b]$). Then $K(X)$ is also $\log$-concave, since it is the convolution of two $\log$-concave functions, $\psi$ and $W$, and the convolution of two $\log$-concave functions is $\log$-concave. Biologically this is quite informative: if the distribution of fitnesses introduced by mutation is $\log$ concave, then the substitution rate is not only decreasing as fitness ($X$) increases, but it is decreasing at least exponentially.

Another natural question is how $K(X)$ relates to $W(Y^{*})$, where $Y^{*}$ is the mean of the fitnesses introduced by mutation. The fact that $w(s)$ is convex tells us immediately that $K(X)\geq W(Y^{*})$, since the expectation of a convex function of a random variable is greater than the function evaluated at the expectation (i.e.~Jensen's inequality). This tells us that the common technique of approximating the distribution of fitnesses introduced by mutation with the mean fitness effect of those mutations will generally result in an underestimate of the substitution rate, at least in the large $N$, fixed $Ns$ regime.

While our identities for the probability of fixation can provide qualitative insight into the rate of evolution, our approximations can produce
quantitative advances by allowing direct evaluation of the necessary integrals. For instance, $K(X)$ has often been evaluated for particular choices of $\psi$. One frequent choice is to set $X=0$ and let $\psi$ be a reflected gamma distribution~\citep[e.g.][]{Kimura79}. Exact evaluation of the resulting integral is possible in terms of zeta functions (or the equivalent infinite series) which then require further approximation. Can we use our approximations for the probability of fixation to get a more useful result? For a gamma distribution with mean $\mu$ and shape $k$, we have
\begin{align}
K(0) &=\int_{-\infty}^{0} \frac{(-S)^{k-1}e^{(k/\mu)S}}{\Gamma(k)(\mu/k)^{k}} W(S)\,dS \\
 &\approx \int_{-\infty}^{0} \frac{(-S)^{k-1}e^{(k/\mu)S}}{\Gamma(k)(\mu/k)^{k}} \left(-Se^{S}+e^{3S/2} \right)\,dS\\
 &=\mu\left(\frac{k}{k+\mu}\right)^{k+1}+\left(\frac{k}{k+3\mu/2}\right)^{k} \label{eq:newgamma}
\end{align}
where we have used the approximation given by Equation~\ref{eq:currentpoonottoS}. Now, we have seen that Equation~\ref{eq:currentpoonottoS} provides an upper bound on the probability of fixation, so that the above expression is an upper bound on the rate of evolution. Furthermore, we have seen that the error in Equation~\ref{eq:currentpoonottoS} as compared to $W(S)$ is no more than $2.4 \%$ and, so our estimate of the substitution rate is no more than $2.4\%$ higher than the value we would have obtained using $W(S)$. Under the additional assumption that $\mu \gg k$ (a common assumption in the literature, since $k$ is typically chosen to be $\leq 1$), we have
\begin{equation}
\label{eq:welch}
K(0)\approx \left(\frac{k}{\mu}\right)^{k}\left(k+(2/3)^{k}\right).
\end{equation} 
This expression is again an upper bound on the substitution rate and turns out to be equivalent to the result in Equation~10 of~\citep{Kimura79}, which was derived using an approximation to the Hurwitz zeta function under the assumption  $\mu \gg k$ (see also the discussion around Equation~23 in~\citealt{Welch08}). For a reflected exponential distribution~\citep[$k=1$, as in][]{Ohta77}, Equation~\ref{eq:welch} reduces further to
\begin{equation}
K(0)\approx \frac{5}{3} \left(\frac{1}{\mu}\right),
\end{equation}
which agrees with Ohta's claim, based on numerical results, that the rate of evolution is inversely proportional to the population size under this model ($-\mu$ is the mean scaled selection coefficient and is therefore proportional to $N$).

One obvious defect of a model that assumes that all mutations are deleterious is that it predicts that the fitness of a population should, over time, decrease indefinitely. One possible solution to this problem is to assume that the genome is made up of an infinite collection of independently evolving biallelic loci such that each fixation of an allele with selection coefficient $S$ results in the creation of a potential mutant with selection coefficient $-S$~\citep[][cf.~\citealt{Bulmer91}]{Piganeau03}. If we suppose that the probability density function of the distribution of mutational effects when all such alleles are fixed at their preferred state is given by $\psi(S)$, then at equilibrium the probability density function of the distribution of mutational effects is given by $\psi(-|S|) /(1+e^{S})$. The most commonly used version of this model chooses $\psi(-|S|)$ to be the pdf of a gamma distribution, in which case the distribution of mutational effects at equilibrium is known as the partially reflected gamma distribution~\citep{Piganeau03,Welch08}. What is surprising is that having analyzed a model of deleterious mutations using the approximation in Equation~\ref{eq:currentpoonottoS}, one can very easily construct an approximation for the rate of evolution under the corresponding ``partially reflected'' model, i.e.~the partially reflected model with the same probability density function $\psi$.

The key to constructing this new approximation is the formula given in Equation~\ref{eq:kcomp}, which tells us that $W(S)=S+W(-S)$. If we think about this as a partitioning of the rate of evolution into one component due to finite population size and another component corresponding to the substitutions that would still occur in an infinite population, we see that the rate of evolution due to finite population size depends only on the distribution of $|S|$ and not on the distribution of $S$ itself. This means that the rate of evolution due to finite population size for a partially reflected model is always the same as the total substitution rate for a model of deleterious mutations with the same distribution of absolute values of selection coefficients. Thus, for the case of a partially reflected gamma distribution, Equation~\ref{eq:newgamma} provides an approximation for the rate of evolution due to finite population size effects (and indeed, an upper bound). 

Now, to find the total rate of evolution, one must of course also find the rate of evolution that would occur in an infinite population. This is given by
\begin{align}
\int_{0}^{\infty} S\,\frac{\psi(-|S|)}{1+e^{S}}\,dS &\approx \int_{0}^{\infty} Se^{-S}\,\psi(-|S|)\,dS \\
 &=\int_{-\infty}^{0} -Se^{S}\,\psi(-|S|)\,dS,
\end{align}
where the approximation is in fact an upper bound. Importantly, the expression in the last line is just the contribution of the term $-Se^{S}$ to the rate of evolution under the corresponding model of deleterious fixations when analyzed using the approximation for the probability of fixation given by Equation~\ref{eq:currentpoonottoS}. Thus, to find the rate of evolution under a partially reflected model of evolution, one can simply take an approximation for the corresponding model of deleterious fixations derived using Equation~\ref{eq:currentpoonottoS} and double the term corresponding to $-Se^{S}$ (e.g.~for the case of a partially reflected gamma distribution, one just doubles the first term in Equation~\ref{eq:newgamma}). Such an approximation is always an upper bound compared to using $W(S)$, and the error is guaranteed to be no more than $14.4\%$ (as can be seen by comparing the contribution to the rate of evolution of a locus with selection coefficient $|S|$ under the approximation, $2|S|e^{-|S|}+e^{-3 |S|/2}$, to the true value, $|S|/\sinh(|S|)$).

Besides ease of computation, such an approximation clarifies the relationship between a model of deleterious mutations and the corresponding partially reflected model. For instance, while it is perhaps obvious that the rate of evolution under a partially reflected distribution is always greater than under the corresponding model of deleterious mutations, the above approximation makes it clear that the increase in the rate of evolution is no more than approximately two-fold (more precisely, the increase can be no more than $2.024$-fold). Furthermore, the approximation makes it clear that this limit is achieved when most selection coefficients are of relatively large magnitude (e.g. $|S|>3$, so that $|S|e^{-|S|} \gg e^{-3 |S|/2}$).

Thus, in summary, the advantage of using the methods described here is not only the ability to derive novel results such as Equation~\ref{eq:newgamma} or our treatment of partially reflected models, but to do so using elementary techniques and with additional guarantees on the accuracy of the approximation.

\section{Discussion}
\label{sec:discussion}

The probability of fixation for a new mutation plays a central role in evolutionary genetics. It quantifies the balance between natural selection and genetic drift in
simple models, and
it forms a key component in more complex models of evolution. Here, we have presented a comprehensive analysis of the fixation probability of
a new mutation, under the Moran process.
We have seen that the logarithm of the fixation probability behaves in a simple manner and have leveraged this simplicity to develop a series of
identities and inequalities (we have also shown how these results can be extended to the situation when an allele is initially present in more than a single copy, see \ref{sec:polymorphic}). Finally, we have introduced a number of new approximations for the fixation probability of a new mutation and summarized their behavior
relative to existing approximations.

Our analysis is based on the remarkable and well-known relation $u_{N}(s)/u_{N}(-s)=e^{(N-1)s}$. This relation suggests a natural and
unified approach towards studying the fixation probability. First, it immediately implies that the logarithm of the fixation probability
behaves in a simple way, which led to our results on $\log$-concavity and related inequalities. Second, it provides a method for deriving
approximations for the fixation probabilty that preserve the characteristics of the long-term dynamics when used in sequential fixations models of mutation-limited
evolution (e.g.~reversibility). The development of such approximations is important because many standard approximations, such as $\max(s,0)$,
produce models that can be grossly inaccurate for long-term prediction~\citep[e.g.~strong-selection weak-mutation models,][]{Gillespie83,Orr05}. 
For instance, whereas the true sequential fixations Markov chain is typically reversible and ergodic, the standard strong-selection approximation
either has absorbing states corresponding to local fitness maxima or else it predicts that fitness increases indefinitely.

Throughout our presentation we have placed emphasis on deriving controlled approximations, that is, approximations where one can bound the magnitude or
direction of the error. Such approximations are important for two reasons. First, they provide peace of mind. One typically uses an approximation when
a certain degree of inaccuracy is acceptable; a controlled approximation guarantees that the inaccuracy is not too great. Second, and perhaps more
importantly, a controlled approximation provides an inequality, and it can therefore be used to prove exact results. 

While our analysis here has been focused on the probability of fixation for a Moran process, many of our results can be extended to address the probability of fixation under the Wright-Fisher process. There are two ways to conduct such an extension, depending on whether one prefers to define $s$ as the difference in the $\log$ fitnesses of the invading and resident types, as we have here, or whether one prefers to use the more traditional definition of $s$ as the ratio of the invading and resident fitnesses minus 1. Keeping with our convention here, one can simply substitute $2s$ for $s$ in any of the expressions that we have derived for a haploid Wright-Fisher model, and, in addition, substitute $2N$ for $N$ in the case of a diploid Wright-Fisher model where fitness is multiplicative within loci. \citet[][supporting text]{Sella05} have shown numerically that the resulting approximation for the probability of fixation has accuracy comparable to the~\citet{Kimura57,Kimura62} expression based on the traditional definition of the selection coefficient; they also provide several conceptual arguments for the superiority of this choice of selection coefficient. However, if one is interested in controlled approximations then the traditional choice of selection coefficient may be superior for the Wright-Fisher process. This is because the~\citet{Kimura57,Kimura62} expression with the traditional definition of the selection coefficient provides an upper bound on the true probability of fixation for (1) the haploid Wright-Fisher process~\citep{Moran60} and (2) the diploid Wright-Fisher process where fitness is additive within loci~\citep{Burger95}. In particular, any of our upper bounds for the probability of fixation under the Moran process can be converted into upper bounds for the Wright-Fisher process by formally substituting $2s$ for $s$  and then treating $s$ as the traditional selection coefficient (while perhaps also substituting $2N$ for $N$ if a diploid model is required).

One result worth discussing in more detail is the simple identity $w(s)=w(-s)+s$. What this identity tells us is that the probability of
fixation of a mutation with selection coefficient $s$ can be partitioned into two components. The first of these components is $\max(s,0)$, which is
the probability of fixation for a new mutation with selection coefficient $s$ in an infinite population (the strong-selection limit). The second of
these components is $w(-|s|)$, which is symmetric around $s=0$ and which captures the effects of finite population size on the probability of fixation. 
This decomposition of $w(s)$ provides insight into many topics that have previously been discussed in the literature. For instance, the symmetry of
finite population size effects around $Ns=0$ provides an additional rationale for why near-neutrality should be defined in terms of the absolute value
of $Ns$, such as (translating to the current context) $|Ns|\ll 2$~\citep{Kimura68b} or $|Ns|<4$~\citep[][cf.~\citealt{Nei05} and~\citealt{Razeto14}]{Li78}. At the same time, the smooth
decay of $w(-|s|)$ makes clear that any specific cutoff is arbitrary. The symmetry of the finite population-size effects around $s=0$ also complements the classical result that the expected waiting time between the introduction of a new mutation destined to fix and the time that it reaches fixation depends only on the magnitude of $Ns$ and not on its sign~\citep[][see also, \citealt{Ewens04}, pp.~170, 188--191]{Maruyama74b,Maruyama74,Taylor06}.

A natural application of this decomposition for $w(s)$ is to ask what proportion of the total substitution rate is due to finite population-size
effects and what proportion would still occur in an infinite population (or, more precisely, what proportion would occur in an infinite population that found
itself fixed for the same set of states as the finite population). Indeed, in unpublished work~\citet{Razeto14} has recently suggested this partition
of the substitution rate to help resolve the neutralist-selectionist debates, where the neutralist position is identified with the claim that most
substitutions are due to the effects of finite population size and the selectionist position with the claim that most substitutions would still occur
in an infinite population; our identity on the symmetry of the finite population effects makes this resolution appear even more natural. Notably, our
results in Section~\ref{sec:applications} show that at equilibrium under any model of independently evolving biallelic loci with an arbitrary
distribution of selection coefficients and no mutational bias, at most $50.6\%$ of substitutions can be due to selection. Thus, any model producing
support for the selectionist position must rely on complications such as non-equilibrial dynamics, mutational biases, epistasis, etc.

This identity for $w(s)$ also provides additional insight into the relationship between the strong-selection weak-mutation approximation and the actual dynamics of evolution under weak mutation. In particular, suppose we have a sequential fixations model with rate matrix $Q$, where $Q(i,j)$ describes the rate at which a population currently fixed for allele $i$ would become fixed for some other allele $j$ and the diagonal entries are chosen so that the rows sum to zero. Then our result shows that we can write $Q=Q_{\operatorname{Sel}}+Q_{\operatorname{Neut}}$ where $Q_{\operatorname{Sel}}$ is the rate matrix for the corresponding strong-selection weak-mutation Markov chain and $Q_{\operatorname{Neut}}$ summarizes the effects of finite population size on the evolutionary dynamics. Indeed, while $Q_{\operatorname{Sel}}$ tends to push populations to higher and higher fitnesses, $Q_{\operatorname{Neut}}$ describes the orthogonal tendency of populations to move laterally across networks of genotypes with similar fitnesses~\citep{Conrad90,Huynen96,Gavrilets97b,AWagner11}. 

On the other hand, it is important to note that this decomposition of the rate matrix $Q$ applies only to the infinitesimal rates and not to the
long-term dynamics.  For instance, $Q_{\operatorname{Sel}}$ will generally define an absorbing, rather than ergodic, Markov chain, and
$Q_{\operatorname{Neut}}$ will typically define a Markov chain whose equilibrium distribution is the same as the distribution of evolution under mutation alone.
However, the equilibrium distribution of the chain defined by $Q$ will not be a simple function of these two distributions. Moreover, the transition matrix for
the process under $Q$, $P_{t}=e^{Qt}$, will generally not equal $e^{Q_{\operatorname{Sel}}t}e^{Q_{\operatorname{Neut}}t}$ or
$e^{Q_{\operatorname{Neut}}t}e^{Q_{\operatorname{Sel}}t}$. Thus, even if we can decompose the instantaneous rate of evolution into the rate due to
selection and the rate due to drift, the long-term effects of selection and drift in finite populations are inextricably intertwined.

\section*{Acknowledgements} We thank Pablo Razeto-Barry for sharing his unpublished manuscript and Warren Ewens for comments. J.B.P. acknowledges funding from the Burroughs Wellcome Fund, the 
David and Lucile Packard Foundation, the James S.~McDonnell Foundation, the Alfred P.~Sloan Foundation, the U.S.~Department of the
Interior (D12AP00025), and the Foundation Question in Evolutionary Biology Fund (RFP-12-16). D.M.M., J.B.P. and C.L.E. acknowledge funding
from the U.S.~Army Research Office (W911NF-12-1-0552). 

\pagebreak
\clearpage

\bibliographystyle{evolution}
\bibliography{MainBibtexDatabase} 

\pagebreak

\appendix
\def\thesection{Appendix \arabic{section}}
\renewcommand{\theequation}{A\arabic{equation}}
\setcounter{equation}{0}

\section{Results for the probability of fixation of an allele segregating at intermediate frequency}
\label{sec:polymorphic}

Almost all of the identities and inequalities presented in the main text carry over to the case of an allele that is initially at frequency greater than $1/N$. To see why, consider the probability of fixation of an allele that begins with $i$ copies in a population of size $N$ under a Moran process:
\begin{equation}
u_{N,i}(s) = \frac{1-e^{-i\,s}}{1-e^{-Ns}}.
\end{equation}
Now, $u_{N,i}(s)$ satisfies the identity
\begin{equation}
\frac{u_{N,i}(s)}{u_{N,i}(-s)}=e^{(N-i)s}, 
\end{equation}
which is exactly analogous to Equation~\ref{eq:wellknown}. Because Equation~\ref{eq:wellknown} underlies many aspects of our analysis, it is not altogether surprising that our results extend to this more general setting.

First, let us consider the generalizations of the identities in Table~\ref{tab:identities}. These are given in Table~\ref{tab:polyidentities}. Second, let us consider the behavior of $\log u_{N,i}(s)$. We have:
\begin{equation}
\frac{\deriv}{\deriv s} \log u_{N,i}(s) = \frac{N}{1-e^{Ns}}-\frac{i}{1-e^{is}}.
\end{equation}
and
\begin{equation}
\frac{\deriv^{2}}{\deriv s^{2}} \log u_{N,i}(s)
=\frac{N^{2}\,u_{N,i}(s)\,u_{N,i}(-s)-i^{2}}{e^{i\,s}+e^{-i\,s}-2}.
\end{equation}
Note that $u_{N,i}(s)\,u_{N,i}(-s)\leq (i/N)^{2}$ (this is easiest to show by analyzing Equation~\ref{eq:appendixhyp}) so that $u_{N,i}(s)$ is again $\log$-concave and the slope of $\log u_{N,i}(s)$ decreases monotonically from $N-i$ to $0$ as $s$ goes from $-\infty$ to $\infty$ in a manner that is symmetric around $s=0$.

One can use this $\log$-concavity, together with geometric arguments analogous to those in Figure~\ref{fig:inequalities}, to extend the inequalities presented in Section~\ref{sec:inequalities} to the situation where an allele begins at intermediate frequency. In particular we have
\begin{equation}
u_{N,i}(s)\leq u_{N,i}(\tilde{s})\,e^{\frac{u'_{N,i}(\tilde{s})}{u_{N,i}(\tilde{s})}(s-\tilde{s})}.
\end{equation}
for all $s$, $\tilde{s}$, and
\begin{equation}
\frac{u_{N,i}(s+c)}{u_{N,i}(\tilde{s}+c)}<\frac{u_{N,i}(s)}{u_{N,i}(\tilde{s})}<e^{(N-i)(s-\tilde{s})}
\label{eq:apineq1}
\end{equation}
for $s>\tilde{s}$, $c>0$, and the direction of the inequalities in Equations~\ref{eq:apineq1} are reversed for $s<\tilde{s}$.

\setid

\begin{table}[p]
\begin{align}
\label{eq:const} &u_{N,i}(s)=1-u_{N,N-i}(s) & & \text{Self-consistency}\\
 &\frac{u_{N,i}(s)}{u_{N,i}(-s)} =e^{(N-i)s}  & &   \text{Reversibility} \\
&\frac{1}{u_{N,i}(s)} = \sum_{k=0}^{N/i-1}e^{-k\,i\,s}  & &   \text{Finite geometric sum (valid if $i$ divides $N$)} \\
 &u_{N,i}(s) =\frac{u_{N-1,i}(s)}{e^{-s}\left(1- u_{N-1,i-1}(s) \right)+u_{N-1,i}(s)} & & \text{Recursive formula} \\
&u_{N,i}(s)-(1-e^{-i\,s})=e^{-i\,s}\,u_{N,i}(-s) & & \text{Comparison with $Ns \gg1$ limit}\\
&u_{N,i}(s) =e^{(N-i)s/2}\left(\frac{\sinh (is/2)}{\sinh (Ns/2)}\right) & &  \text{Hyperbolic identity} \\
 \label{eq:appendixhyp}&u_{N,i}(s)\,u_{N,i}(-s) =\left(\frac{\sinh (is/2)}{\sinh (Ns/2)}\right)^{2} & & \text{Hyperbolic identity II} \\
 & u_{N,i}'(s) =  i\left(1-u_{N,i}(s)\right)-e^{-is}u'_{N,i}(-s) & & \text{Derivative with respect to $s$} 
\end{align}
\caption{\label{tab:polyidentities} Extension of identities in Table~\ref{tab:identities} to initial frequencies greater than $1/N$.} 
\end{table}
\afterpage{\clearpage}

\setreg

\end{document}